\documentclass{elsart}
\usepackage{latexsym}
\usepackage{stmaryrd}

\usepackage[dvips]{graphicx}

\usepackage{times}
\usepackage[T1]{fontenc}
\normalfont
\usepackage{mathptm}

\def\linearimp{\mathop{- \hspace{-.025in} \circ}}
\def\linearequiv{\mathop{\circ \hspace{-.055in} - \hspace{-.055in} \circ}}
\newcommand{\LIMP}{\linearimp}
\newcommand{\LQUIV} 
{\linearequiv}

\newcommand{\PAR}{\bindnasrepma}
\newcommand{\WITH}{\binampersand}
\newcommand{\PLUS}{\oplus}
\newcommand{\TENS}{\otimes}

\def\depth{\mathop{\rm depth} \nolimits}
\def\max{\mathop{\rm max} \nolimits}
\def\order{\mathop{\rm order} \nolimits}
\def\measure{\mathop{\rm measure} \nolimits}

\def\arity{\mathop{\rm arity} \nolimits}

\def\abb{\mathop{\rm abb} \nolimits}
\def\subabb{\mathop{\rm sabb} \nolimits}

\newtheorem{lemma}{Lemma}
\newtheorem{theorem}{Theorem}
\newtheorem{definition}{Definition}
\newtheorem{corollary}{Corollary}
\newtheorem{proposition}{Proposition}
\newtheorem{example}{Example}

\newenvironment{remark}{\begin{flushleft}{\it Remark.} \ \ }{\end{flushleft}}
\newenvironment{proof}{\begin{flushleft}{\it Proof.} \ \ }{\end{flushleft}}

\begin{document}
\begin{frontmatter}
\title{Weak Typed B\"{o}hm Theorem on IMLL}
\author{Satoshi Matsuoka}
\address{
National Institute of Advanced Industrial Science and Technology,
1-1-1 Umezono,
Tsukuba, Ibaraki,
305-8561 Japan}
\ead{matsuoka@ni.aist.go.jp}

\begin{abstract}
In the B\"{o}hm theorem workshop on Crete island, Zoran Petric called Statman's 
``Typical Ambiguity theorem'' {\it typed B\"{o}hm theorem}. Moreover, he gave 
a new proof of the theorem
based on set-theoretical models of the simply typed lambda calculus. \\
In this paper, we study the linear version of the typed B\"{o}hm theorem 
on a fragment of Intuitionistic Linear Logic.
We show that 
in 
the multiplicative fragment of intuitionistic linear logic 
without the multiplicative unit ${\bf 1}$ (for short IMLL) weak typed B\"{o}hm theorem holds. 
The system IMLL exactly corresponds to the linear lambda calculus 
without exponentials, additives and logical constants.
The system IMLL also exactly corresponds to the free 
symmetric monoidal closed category without the unit object. 
As far as we know, our separation result is the first one 
with regard to these systems in a purely syntactical manner. 
\end{abstract}
\end{frontmatter}

\section{Introduction} \label{Introduction}
In \cite{DP01}, Dosen and Petric called Statman's 
``Typical Ambiguity theorem''~\cite{Sta83} {\it typed B\"{o}hm theorem}. Moreover, they gave 
a new proof of the theorem
based on set-theoretical models of the simply typed lambda calculus. \\
In this paper, we study the linear version of the typed B\"{o}hm theorem 
on intuitionistic multiplicative Linear Logic without the multiplicative unit
${\bf 1}$ (for short IMLL).
We consider the typed version of the following statement:
\begin{quote}
There are two different closed $\beta\eta$-normal terms $\underbar{0}$ and $\underbar{1}$ such that 
if $s$ and $t$ are closed untyped normal $\lambda$-terms, 
and $s \neq_{\beta \eta} t$ then,
there is a context $C[]$ such that\\
                \[ C[s] =_{\beta\eta} \underbar{0} \, \, \, \mbox{ and } \, \, \, 
                 C[t] =_{\beta\eta} \underbar{1} \] 
\end{quote}
We call the statement {\it weak untyped B\"{o}hm theorem}.
In this paper, we show that 
the typed version of weak B\"{o}hm theorem holds in IMLL.\\
The theorem is nontrivial because 
the system IMLL is rather weak in expressibility. 
Hence, a careful analysis on IMLL proof nets is needed.
The system IMLL exactly corresponds to the linear lambda calculus 
without exponentials, additives and logical constants.
A version of the linear lambda calculus can be found in \cite{MO03}.
The system IMLL also exactly corresponds to the free 
symmetric monoidal closed category without the unit object(see \cite{MO03}).
As far as we know, the result we prove in this paper is the first one
with regard to these systems in a purely syntactical manner.\\
On the other hand, we call the following statement {\it strong untyped B\"{o}hm theorem}:
\begin{quote}
For any untyped $\lambda$-terms $a$ and $b$,  
if $s$ and $t$ are closed untyped normal $\lambda$-terms, 
and $s \neq_{\beta \eta} t$ then,
there is a context $C[]$ such that\\
                \[ C[s] =_{\beta\eta} a \, \, \, \mbox{ and } \, \, \, 
                 C[t] =_{\beta\eta} b \] 
\end{quote}
We could not prove the typed version of the statement in the system IMLL.
But so far we proved the typed version of the statement w.r.t a very limited fragment including additive connectives of Linear Logic (see Section~\ref{CONC}). Also note that 
the weak statement and the strong statement are trivially equivalent in the untyped $\lambda K$-calculus (i.e., the usual $\lambda$-calculus) and in the simply typed $\lambda$-calculus (if type instantiation is allowed) because 
both systems allow unrestricted weakening. 
\\
Although currently we have not developed applications of the theorem,
Statman's typical ambiguity theorem has several applications in foundations of programming languages (for example \cite{SP00}). Intuitionistic Linear Logic has become more important because 
game semantics is successful as a method giving fully abstract semantics for many programming languages
and Intuitionistic Linear Logic can be seen as a foundation for game semantics. 
We hope that our result contributes to further analysis of proofs and further applications on Linear Logic.\\
{\it Related works} \ Our work is obviously based on that of \cite{Sta83} (see also \\
\cite{Sta80,Sta82,SD92}). 
As we said before, however, our result can not be derived directly from that of 
\cite{Sta83}, mainly because of lack of unrestricted weakening in IMLL. 
It is also interesting that  unlike ours, the separability result of \cite{Sta83} cannot be 
obtained simply by substituting a type which has only two closed normal terms: a type which should be instantiated depends on the maximal number of occurrences of variables if you want to restrict the type to have only a finite number of closed terms, since the simply typed lambda calculus allows unrestricted contraction.
Of course, you can choose a type which has infinitely many closed terms like
the Church integer. But IMLL does not have such a type. \\
On the other hand, recently, some works \cite{DP00,Jol00,TdF00,TdF03,LT04} other than \cite{DP01} have been also done 
on similar topics to typed B\"{o}hm theorem. 
However, the system with which \cite{Jol00} and \cite{DP00,DP01} dealt is the simply typed lambda calculus or the free cartesian closed category, not  
IMLL. The works of \cite{TdF00,TdF03,LT04} are technically completely different from ours. \\
{\it The structure of the paper} \ Section~\ref{IMLL} and \ref{EQUALITY} give a definition of IMLL proof nets and an equality on them.
Section~\ref{THIRDRED} and \ref{VALSEP} give a proof of weak typed B\"{o}hm theorem on the implicational fragment of IMLL
(for short IIMLL). 
Section~\ref{EXIMLL} describes a reduction of an unequation of IMLL proof nets to that of IIMLL proof nets.
By the reduction we complete a proof of weak typed B\"{o}hm theorem on IMLL.
Section~\ref{CONC} discusses extensions of our result to IMLL with the multiplicative constant {\bf 1}, MLL, and IMLL with additives.
\section{The IMLL systems} \label{IMLL}
In this section, we present intuitionistic multiplicative proof nets.
We also call these {\it IMLL proof nets}.
\begin{definition}(MLL formulas)
MLL formulas (or simply formulas) (F) is inductively constructed from atomic formulas (P) 
and logical connectives:
\begin{itemize}
\item $P = p$ 
\item $F = P \, | \, F \TENS F \, | \, F \PAR F$.
\end{itemize}
In this paper, we only consider MLL formulas with the only one propositional variable $p$. 
All the results in this paper can be easily extended to the general case with 
denumerable propositional variables, since we just substitute $p$ for these
propositional variables.
\end{definition}

\begin{figure}[htbp]
\begin{center}
\includegraphics[scale=.5]{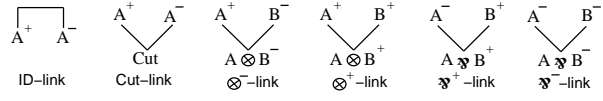}
\caption[the links we use in this paper]{the links we use in this paper}  
\label{imlllink}
\end{center}
\end{figure}

\begin{definition}(IMLL formulas)
An IMLL formula is a pair $\langle A, pl \rangle$
where A is an MLL formula and pl is an element of $\{ +, - \}$, 
where $+$ and $-$ are called Danos-Regnier polarities. 
A formula $\langle A, pl \rangle$ is written as $A^{pl}$. 
A formula with $+$ (resp. $-$) polarity is called $+$-formula or positive formula
(resp. $-$-formula or negative formula).
\end{definition}
Figure~\ref{imlllink} shows the links we use in this paper. 
In Figure~\ref{imlllink}, 
\begin{enumerate}
\item In ID-link, $A^+$ and $A^-$ are called conclusions of the link.
\item In Cut-link, $A^+$ and $A^-$ are called premises of the link.
\item In $\TENS^-$-link (resp. $\PAR^+$-link) $A^+$ (resp. $A^-$) is called the left premise, $B^-$ (resp. $B^+$) the right premise 
and ${A \TENS B}^-$ (resp. ${A \PAR B}^+$) the conclusion of the link.
\item In $\TENS^+$-link (respectively $\PAR^-$-link), $A^+$ (resp. $A^-$) is called the left premise,  $B^+$ (resp. $B^-$) the right premise 
and ${A \TENS B}^+$ (resp. ${A \PAR B}^-$) the conclusion of the link.
\end{enumerate}

\begin{figure}[htbp]
\begin{center}
\includegraphics[scale=.5]{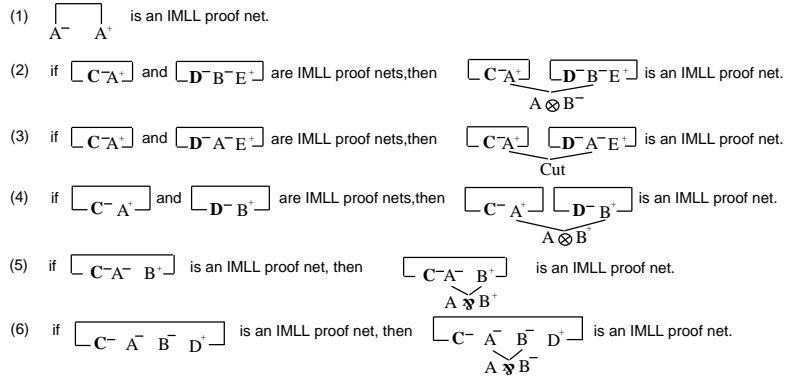}
\caption[the definition of IMLL proof nets]{the definition of IMLL proof nets}  
\label{imllpn}
\end{center}
\end{figure}

Figure~\ref{imllpn} shows that {\it IMLL proof nets} are defined inductively, 
where ${\bf C}^-$ and ${\bf D}^-$ are a list of $-$-formulas.\footnote{An anonymous referee requested to give a correspondence between IMLL proof nets and linear lambda calculus. But the correspondence is a well-known fact (see \cite{MO03}). To do such a thing would just make this paper lengthy unnecessarily. So we refuse the request.}
If $\Theta$ is an IMLL proof net and $\Theta$ is defined without using 
clauses (4) and (6), then we say that $\Theta$ is an IIMLL proof net. 
In the definition of IMLL proof nets, we permit 
'crossings' of links, because the IMLL system has an exchange rule. 
A typical example of such a crossing is that of 
Figure~\ref{imll-pn-ex2}. In an IMLL proof net $\Theta$, a formula occurrence $A$ is a conclusion of
$\Theta$ if $A$ is not a premise of a link. 

\begin{figure}[htbp]
\begin{center}
\includegraphics[scale=.5]{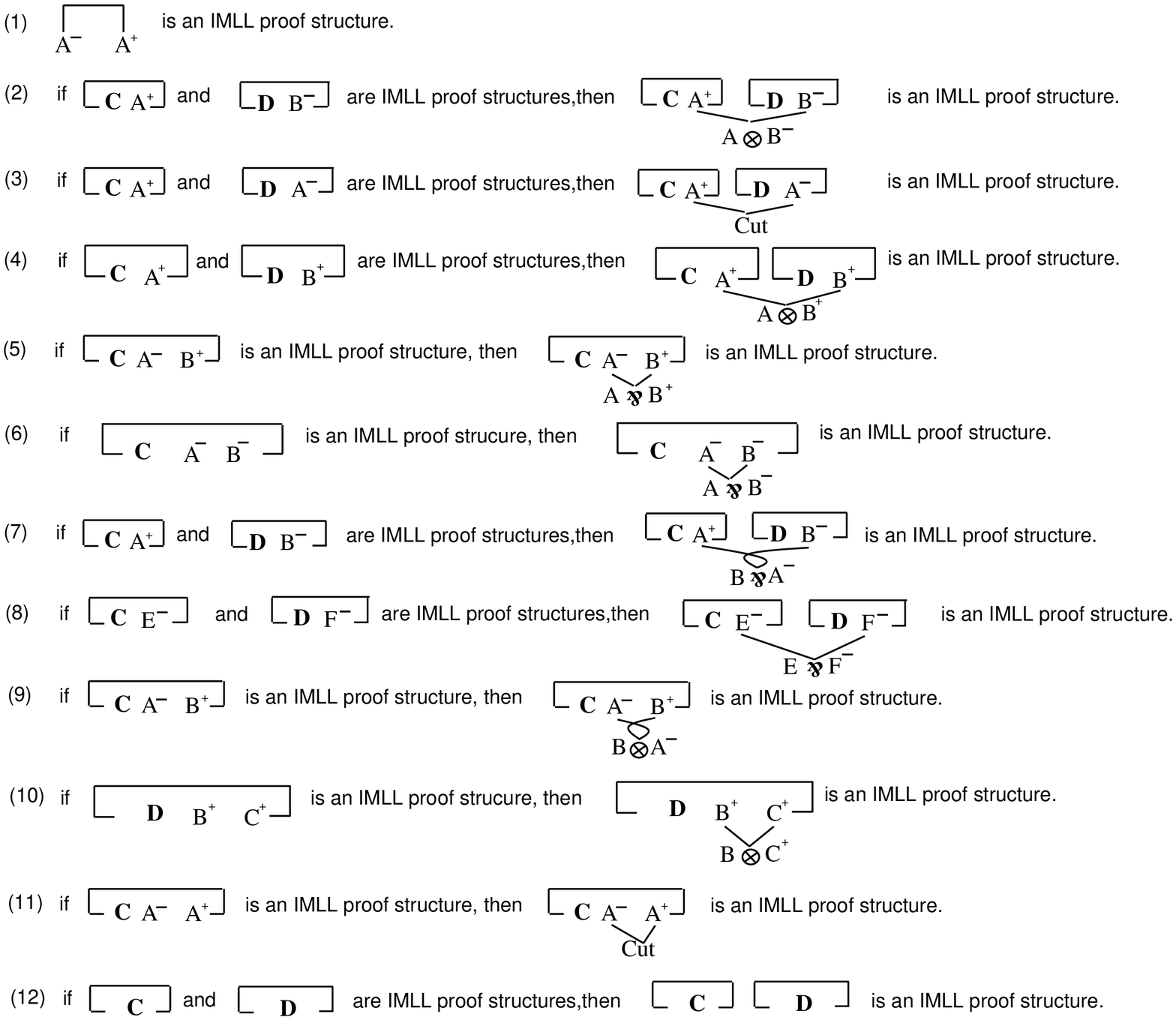}
\caption[the definition of IMLL proof structures]{the definition of IMLL proof structures}  
\label{imllps}
\end{center}
\end{figure}

\begin{figure}[htbp]
\begin{center}
\includegraphics[scale=.5]{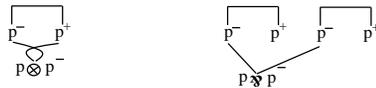}
\caption[two examples of IMLL proof structures]{two examples of IMLL proof structures}  
\label{ps-ex1}
\end{center}
\end{figure}

\begin{figure}[htbp]
\begin{center}
\includegraphics[scale=.5]{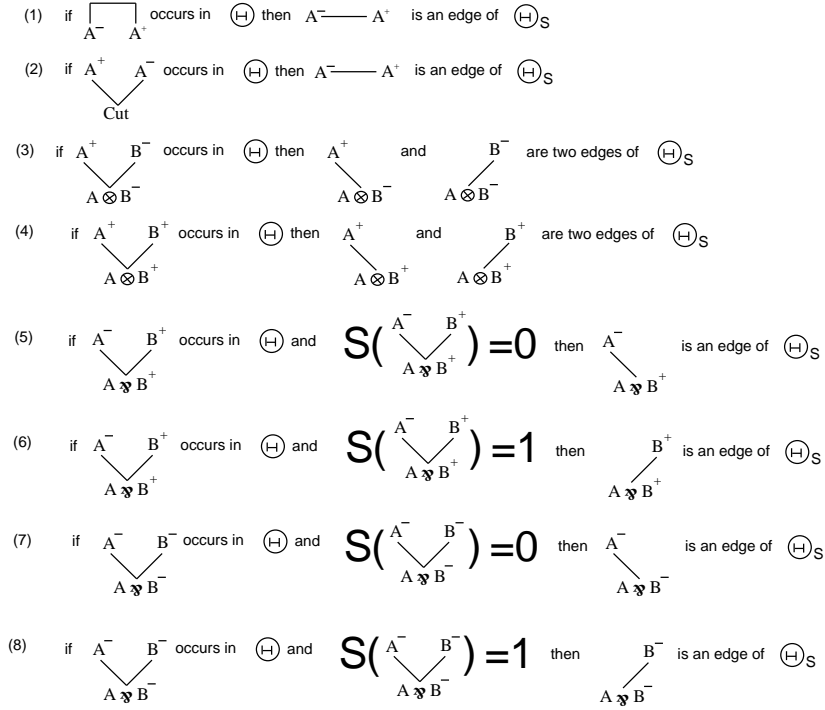}
\caption[the rules for the generation of the edges of a Danos-Regnier graph $\Theta_S$]{the rules for the generation of the edges of a Danos-Regnier graph $\Theta_S$}  
\label{dr-graph-edges}
\end{center}
\end{figure}

Next we give the graph-theoretic characterization of IMLL proof nets, 
following \cite{Gir96}, because we use this in the proof of Lemma~\ref{reductionlemma}. 
The characterization was firstly proved in \cite{Gir87} and 
an improvement was given in \cite{DR89}. 
First we define {\it IMLL proof structures}.
Figure~\ref{imllps} shows that IMLL proof structures are defined inductively,
where {\bf C} and {\bf D} are a list whose element is a $-$-formula or a $+$-formula.
Note that the rules from (1) to (6) can be regarded to be generalized ones of that of 
IMLL proof nets. So, the set of the IMLL proof nets is a subset of 
the set of the IMLL proof structures. 
For example, Figure~\ref{ps-ex1} shows two examples of typical 
IMLL proof structures that are not IMLL proof nets.\\
In order to characterize IMLL proof nets among IMLL proof structures, 
we introduce {\it Danos-Regnier graphs}. 
Let $\Theta$ be an IMLL proof structure. 
We assume that we are given a function $S$ from the set of the occurrences of
$\PAR$-links in $\Theta$ to $\{ 0, 1 \}$. 
Such a function is called a {\it switching function} for $\Theta$.
Then the Danos-Regnier graph $\Theta_S$ for $\Theta$ and $S$ is a 
undirected graph such that
\begin{enumerate}
\item the nodes are all the formula occurrences in $\Theta$, and 
\item the edges are generated by the rules of Figure~\ref{dr-graph-edges}.
\end{enumerate}

\begin{figure}[htbp]
\begin{center}
\includegraphics[scale=.5]{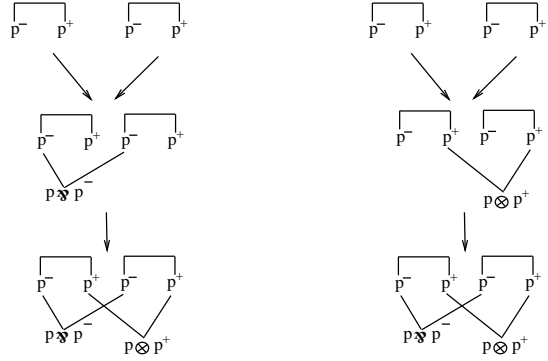}
\caption[an illegal derivation and a legal derivation of the same IMLL proof net]{an illegal derivation and a legal derivation of the same IMLL proof net}  
\label{ps-to-pn-conv-ex1}
\end{center}
\end{figure}

\begin{theorem}[\cite{Gir87} and \cite{DR89}]
\label{seqthm}
An IMLL proof structure $\Theta$ is an IMLL proof net 
iff 
for each switching function $S$ for $\Theta$, 
the Danos-Regnier graph $\Theta_S$ is acyclic and connected.
\end{theorem}
A meaning of the theorem is that 
even though we obtain an IMLL proof structure from an illegal derivation
as a derivation of IMLL proof nets, 
if the proof structure satisfies the criterion of the theorem, 
then we obtain a legal derivation of IMLL proof nets for the IMLL proof structure, i.e., the IMLL proof structure is an IMLL proof net. 
Figure~\ref{ps-to-pn-conv-ex1} shows the situation: 
the left derivation of Figure~\ref{ps-to-pn-conv-ex1} is an illegal 
derivation of IMLL proof nets. But since the derived IMLL proof structure
satisfies the criterion of the theorem, 
the IMLL proof structure is an IMLL proof net and we obtain the right derivation 
of Figure~\ref{ps-to-pn-conv-ex1} for the IMLL proof net.

\begin{figure}[htbp]
\begin{center}
\includegraphics[scale=.5]{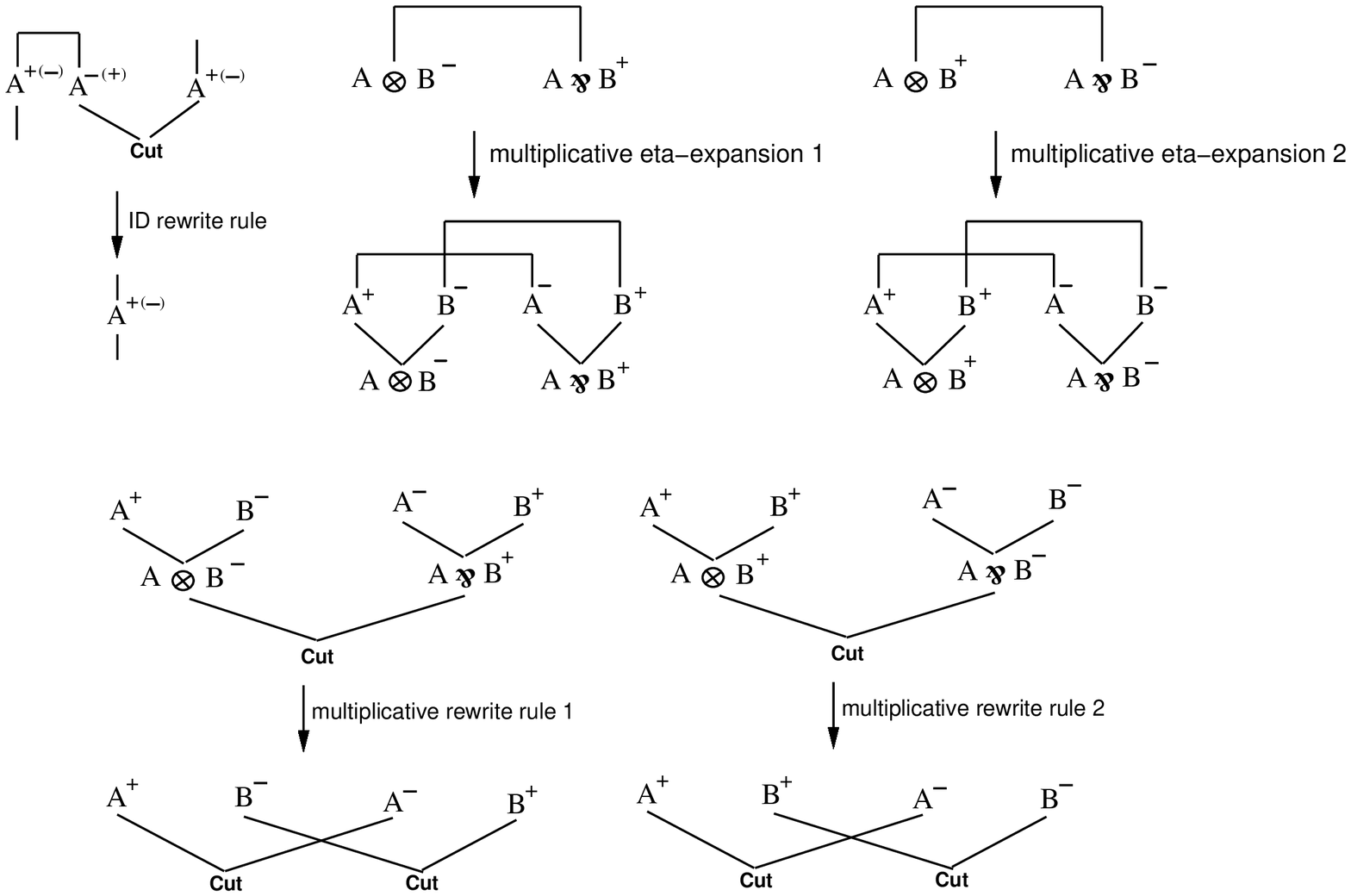}
\caption[the rewrite rules we use in this paper]{the rewrite rules we use in this paper}  
\label{imllCutElim}
\end{center}
\end{figure}

Next we define reduction on IMLL proof nets. 
Figure~\ref{imllCutElim} shows the rewrite rules we use in this paper.
The ID and multiplicative rewrite rules are usual ones.
The multiplicative $\eta$-expansion is the usual $\eta$-expansion in Linear Logic.
We denote the reduction relation defined by these five rewrite rules by $\to^\ast$.  
The one step reduction of $\to^\ast$ is denoted by $\to$.
In the following subsection we show that 
strong normalizability and confluence w.r.t $\to$ holds. 
Hence without mention, we identify an IMLL proof net  
with the normalized net.

\paragraph*{Abbreviations}
In the following we use an abbreviation using linear implication $\LIMP$ 
instead of $\PAR$ in order to relate our IMLL formulas to usual IMLL formulas in the linear lambda calculus (for example, in  \cite{MO03}). 
\begin{enumerate}
\item $\abb(A^+) = {\subabb(A^+)}^+$ \ \ \ \ $\abb(A^-) = {\subabb(A^-)}^-$
\item $\subabb({p}^-) = \subabb({p}^+) = p$ \ \ \ \ 
\item $\subabb({A \TENS B}^-) = \subabb(A^+) \LIMP \subabb(B^-)$ \ \ \ \ 
	$\subabb({A \PAR B}^+) = \subabb(A^-) \LIMP \subabb(B^+)$
\item $\subabb({A \TENS B}^+) = \subabb(A^+) \TENS \subabb(B^+)$ \ \ \ \ 
	$\subabb({A \PAR B}^-) = \subabb(A^-) \TENS \subabb(B^-)$
\end{enumerate}
For example, $\abb({p \PAR (((p \TENS p) \PAR (p \TENS p)) \PAR p)}^+)$ is
${p \LIMP (((p \LIMP p) \TENS (p \LIMP p)) \LIMP p)}^+$.
We identify an IMLL formula $A^\epsilon$ with $\abb(A^\epsilon)$, 
where $\epsilon=+$ or $-$. 
The notation is confusing a little bit: for example, 
$\abb({p \PAR p}^-) = {p \TENS p}^-$.
This is due to the mismatch between the proof-nets notation and 
the linear lambda calculus notation. However, 
from surrounding contexts, i.e., from  whether $\PAR$ or $\LIMP$ is used, we can 
easily judge which notation is adopted. 

\subsection{Strong normalizability and confluence on the IMLL system}
We believe that these two theorems are folklore. 
We just give the following proofs by a request for an anonymous referee.
The strong normalizability is almost trivial. 
The confluence on IMLL is more complicated because 
in the IMLL with the multiplicative $\eta$-expansion 
one-step confluence does not hold unlike the IMLL without the rewrite rule.
But we do not think that the proofs that we give here are difficult to understand.
If you have no doubt about  the strong normalizability and confluence on the IMLL system, 
you can skip this subsection.

\begin{definition}[the SN size of an ID-link and the SN size of a Cut-link]
The SN size of an ID-link is the size of a conclusion, that is, 
the number of the occurrences of logical connectives in the premise.
Note that the choice between a conclusion and 
the other conclusion is indifferent. 
Also note that the SN size of 
an ID-link with two atomic formulas as the conclusions is 0.
The SN size of a Cut-link is the size of a premise plus 1.
With regard to the SN size of a Cut link, the same remark about the choice between a premise and the other premise 
as that of an ID-link is also applied. 
Also note that the SN size of 
a Cut-link with two atomic formulas as the premises is 1.
\end{definition}

\begin{definition}[the SN size of an IMLL proof net]
The SN size of an IMLL proof net $\Theta$ is 
the sum of the SN sizes of all the occurrences of Cut-links and ID-links in
$\Theta$. 
\end{definition}

\begin{proposition}[Strong normalizability on the IMLL system]
\label{SNIMLL}
Let $\Theta$ be an IMLL proof net. $\Theta$ is strong normalizing.
\end{proposition}

\begin{figure}[htbp]
\begin{center}
\includegraphics[scale=.5]{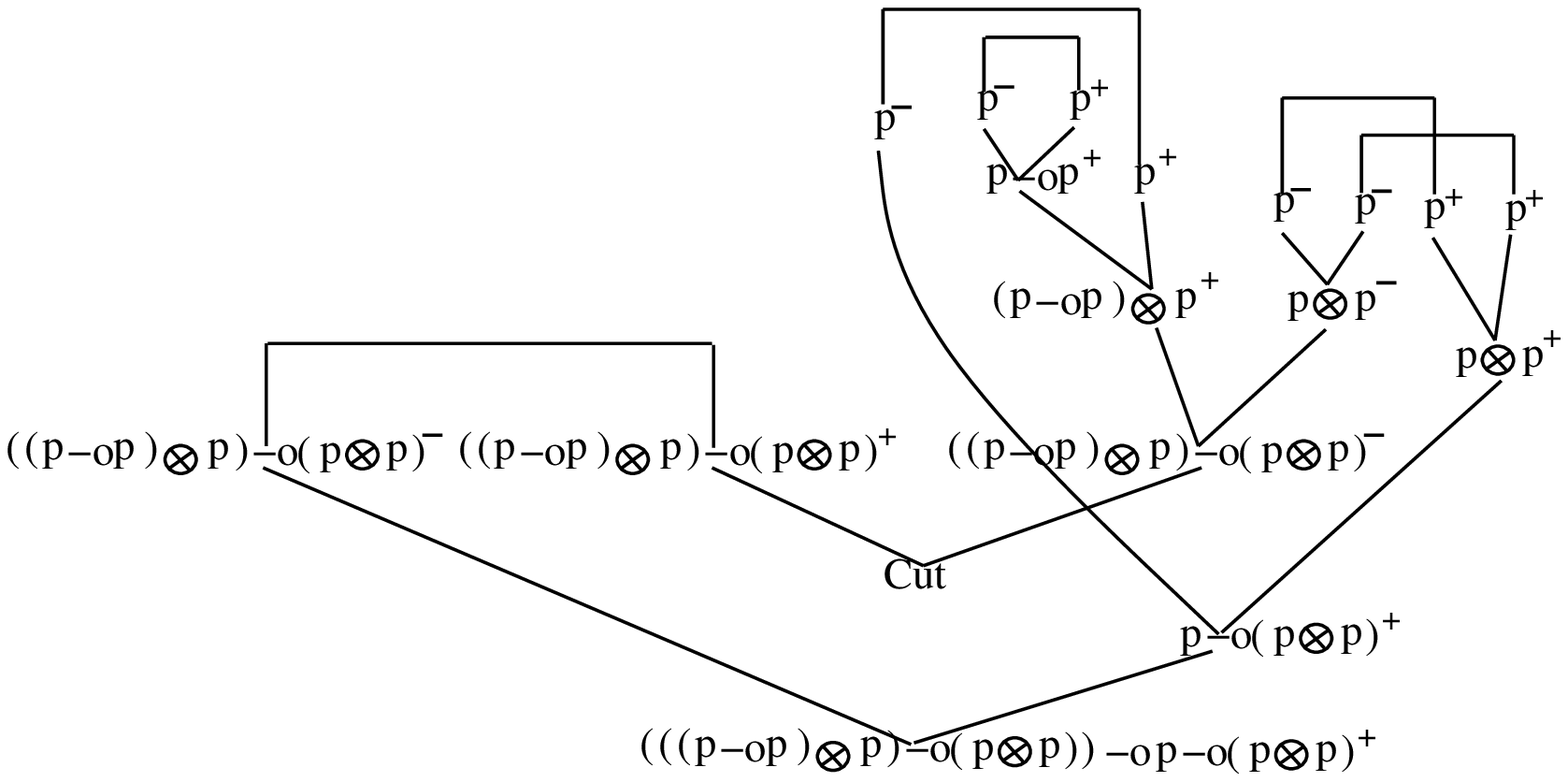}
\caption[an example of IMLL proof nets with Cut-links $\Theta_1$]
{an example of IMLL proof nets with Cut-links $\Theta_1$}
\label{sn-ex1}
\end{center}
\end{figure}

\begin{figure}[htbp]
\begin{center}
\includegraphics[scale=.5]{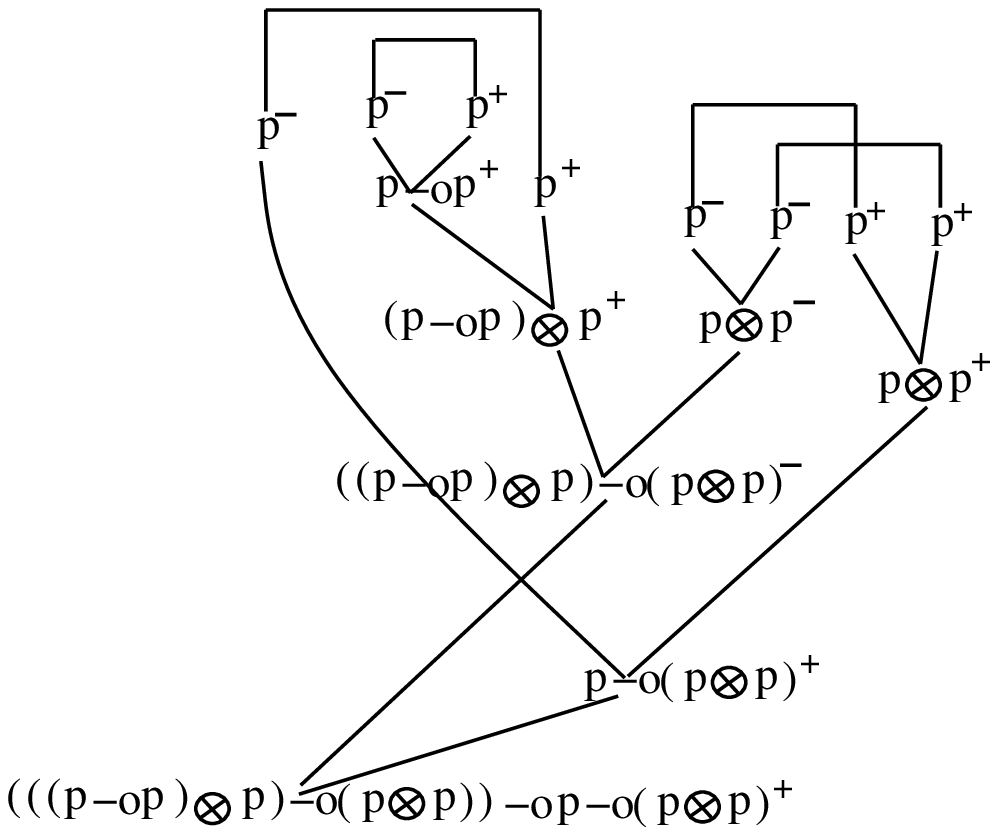}
\caption[the IMLL proof net $\Theta_2$ obtained from $\Theta_1$ by the ID rewrite rule]
{the IMLL proof net $\Theta_2$ obtained from $\Theta_1$ by the ID rewrite rule}  
\label{sn-ex2}
\end{center}
\end{figure}

\begin{figure}[htbp]
\begin{center}
\includegraphics[scale=.5]{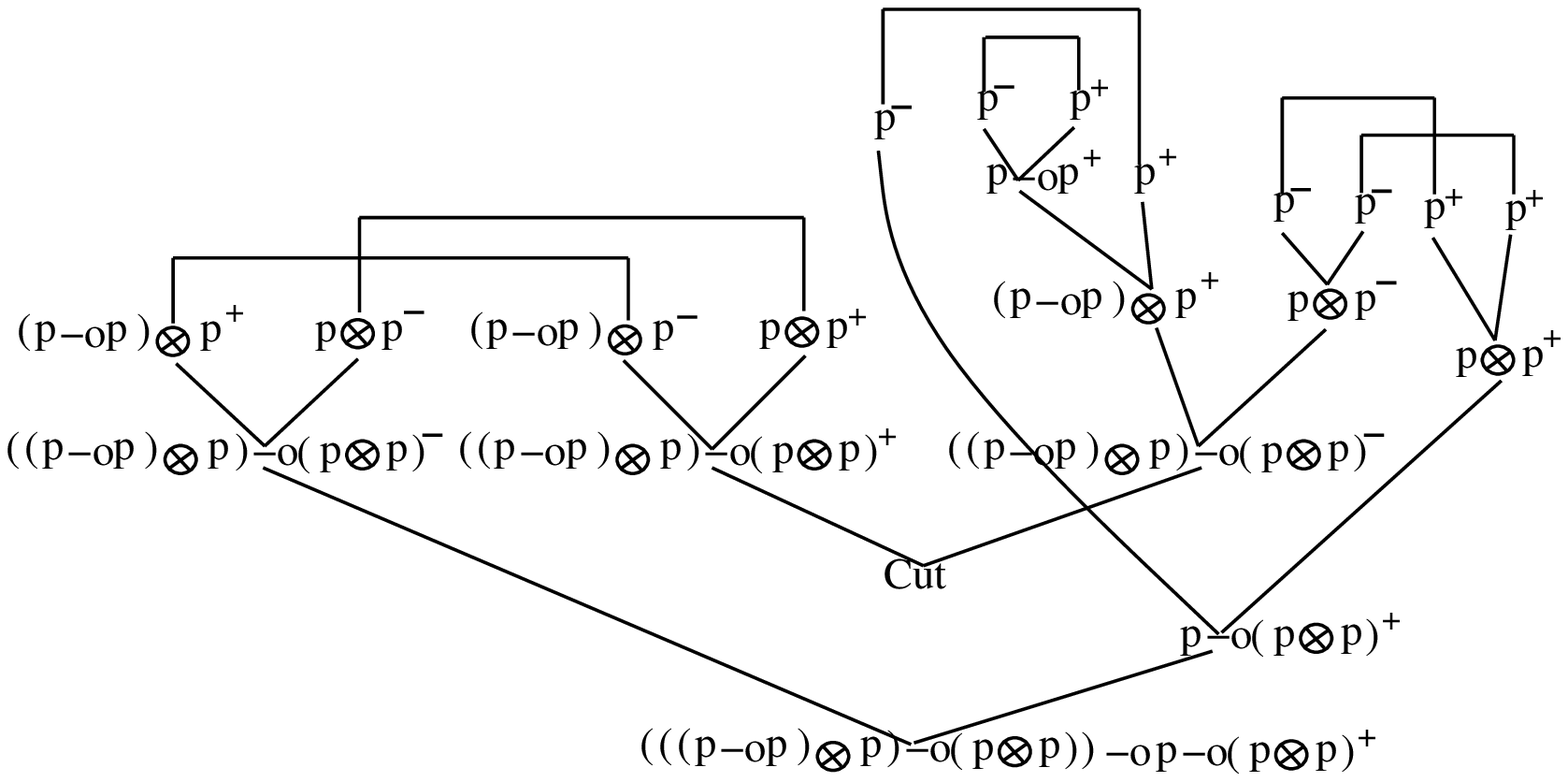}
\caption[the IMLL proof net $\Theta_3$ obtained from $\Theta_1$ by the multiplicative $\eta$-expansion 1]
{the IMLL proof net $\Theta_3$ obtained from $\Theta_1$ by the multiplicative $\eta$-expansion 1}  
\label{sn-ex3}
\end{center}
\end{figure}

\begin{proof}
Let $\Theta \to \Theta'$.
Then in any case where $\Theta$ reduces to $\Theta'$ by a rule in 
Figure~\ref{imllCutElim}, we can easily see the SN size of $\Theta'$ is less 
than that of $\Theta$. 
$\Box$
\end{proof}
For example, the SN size of $\Theta_1$ in Figure~\ref{sn-ex1} is 9.
Then $\Theta_1 \to \Theta_2$ by the ID rewrite rule, 
where $\Theta_2$ is the IMLL proof net of Figure~\ref{sn-ex2}.
The SN size of $\Theta_2$ is 0.
On the other hand $\Theta_1 \to \Theta_3$ by the multiplicative $\eta$-expansion 1, where $\Theta_3$ is the IMLL proof net of Figure~\ref{sn-ex3}. 
The SN size of $\Theta_3$ is 8. 

\begin{figure}[htbp]
\begin{center}
\includegraphics[scale=.5]{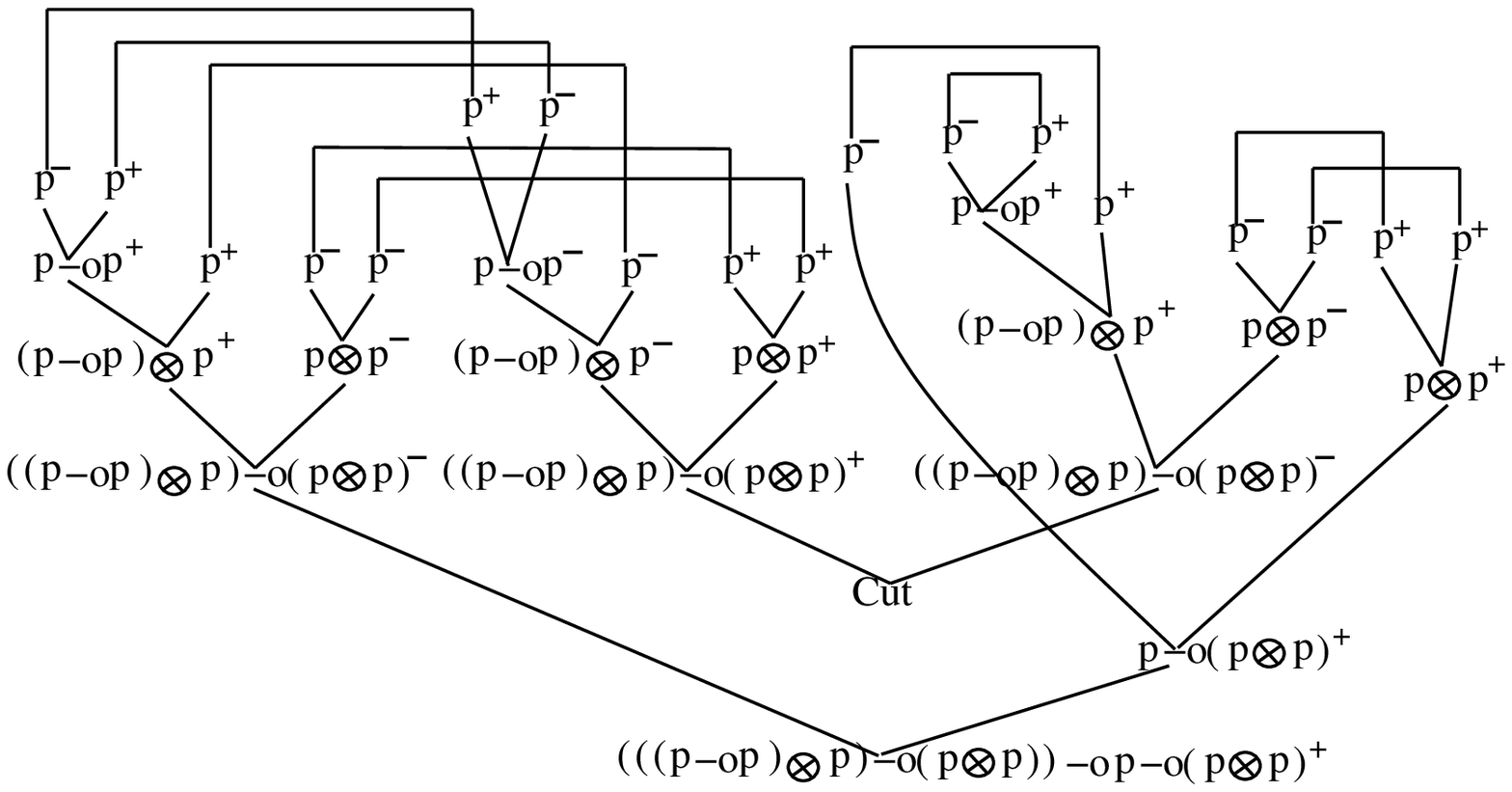}
\caption[the IMLL proof net $\Theta_4$ obtained from $\Theta_3$ by applying the multiplicative $\eta$-expansion three time]
{the IMLL proof net $\Theta_4$ obtained from $\Theta_3$ by applying the multiplicative $\eta$-expansion three time}
\label{sn-ex4}
\end{center}
\end{figure}

\begin{figure}[htbp]
\begin{center}
\includegraphics[scale=.5]{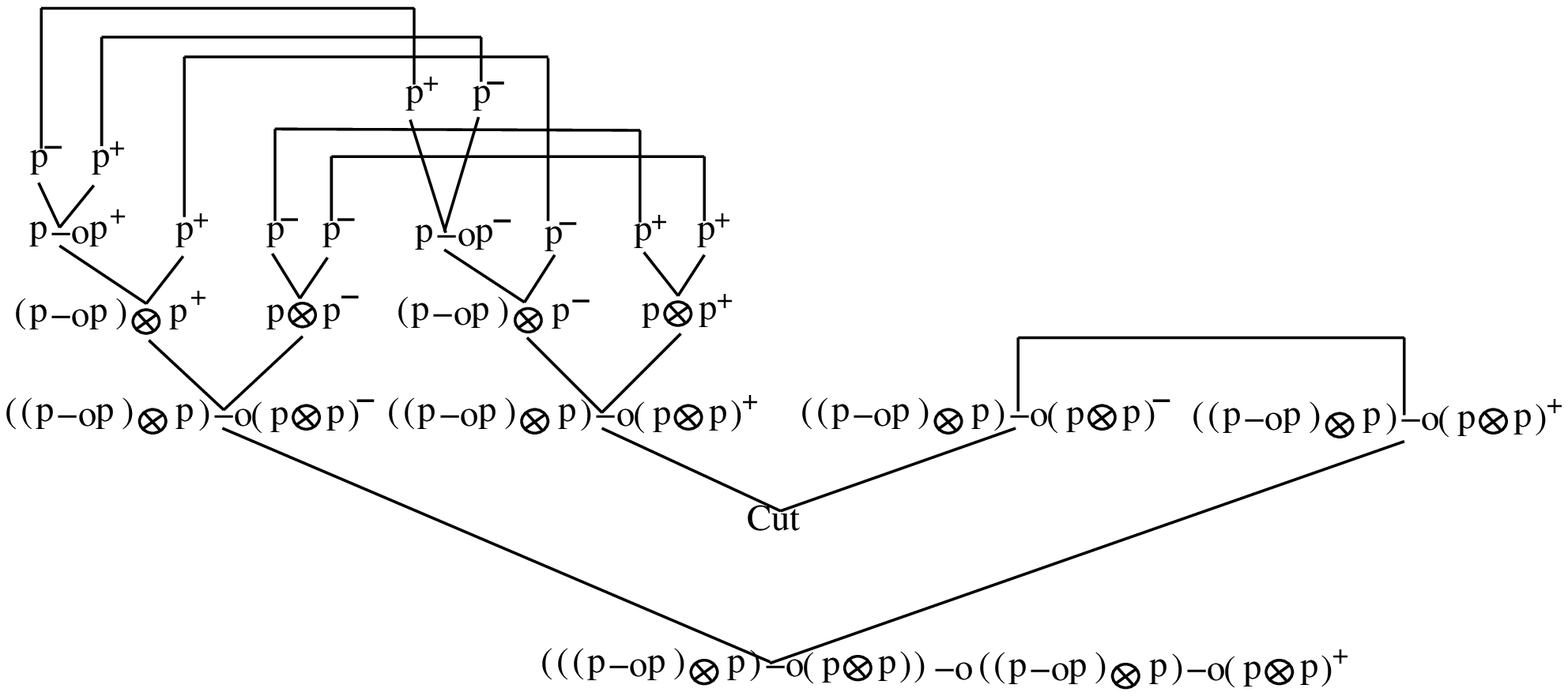}
\caption[another example of IMLL proof nets with Cut-links $\Theta'_1$]
{another example of IMLL proof nets with Cut-links $\Theta'_1$}
\label{sn-ex1-dash}
\end{center}
\end{figure}

\begin{figure}[htbp]
\begin{center}
\includegraphics[scale=.5]{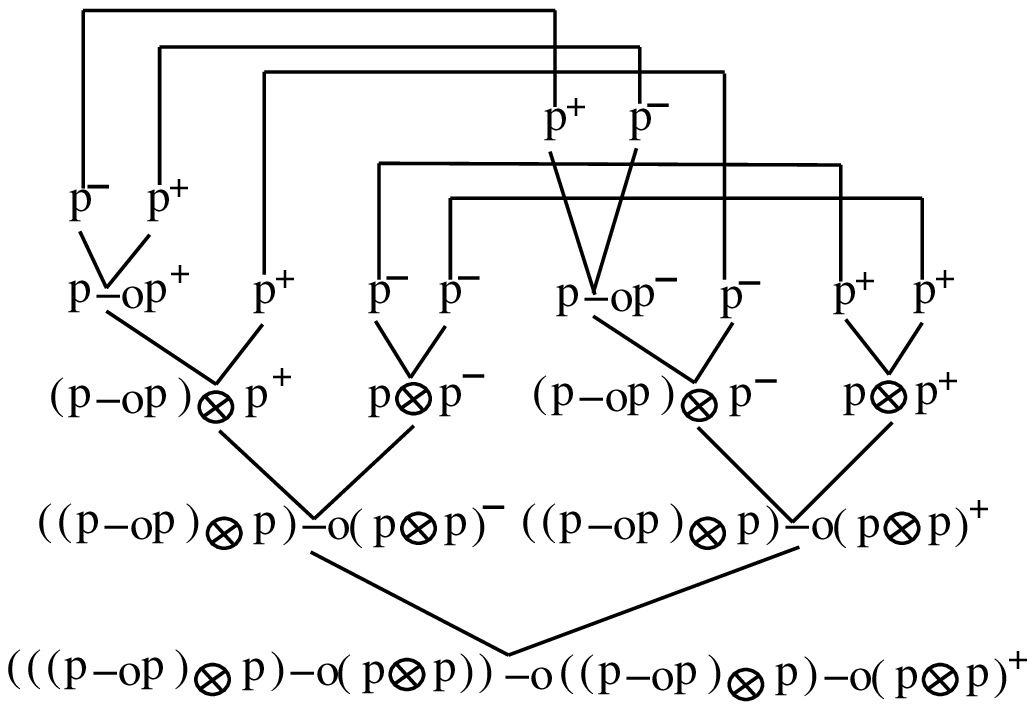}
\caption[the IMLL proof net $\Theta'_2$ obtained from $\Theta'_1$ by the ID rewrite rule]
{the IMLL proof net $\Theta'_2$ obtained from $\Theta'_1$ by the ID rewrite rule}
\label{sn-ex2-dash}
\end{center}
\end{figure}

\begin{figure}[htbp]
\begin{center}
\includegraphics[scale=.5]{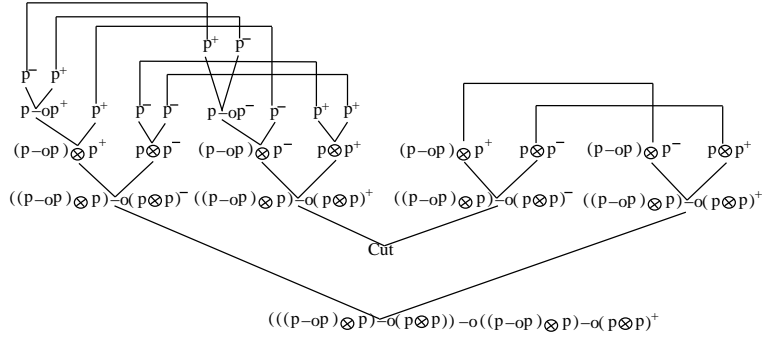}
\caption[the IMLL proof net $\Theta'_3$ obtained from $\Theta'_1$ by the multiplicative $\eta$-expansion 1]
{the IMLL proof net $\Theta'_3$ obtained from $\Theta'_1$ by the multiplicative $\eta$-expansion 1}
\label{sn-ex3-dash}
\end{center}
\end{figure}

\begin{figure}[htbp]
\begin{center}
\includegraphics[scale=.5]{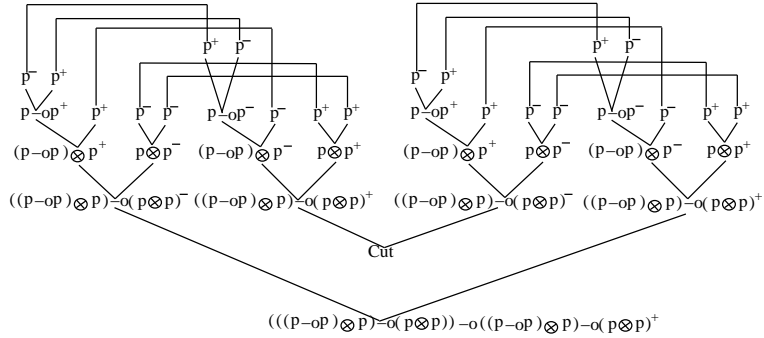}
\caption[the IMLL proof net $\Theta'_4$ obtained from $\Theta'_3$ by applying the multiplicative $\eta$-expansion three time]
{the IMLL proof net $\Theta'_4$ obtained from $\Theta'_3$ by applying the multiplicative $\eta$-expansion three time}
\label{sn-ex4-dash}
\end{center}
\end{figure}

Next, we consider the confluence on the IMLL system. \\
Figure~\ref{sn-ex1}, Figure~\ref{sn-ex2}, and Figure~\ref{sn-ex3} show 
a counterexample of one-step confluence in the IMLL system with the multiplicative $\eta$-expansion, 
since $\Theta_3$ of Figure~\ref{sn-ex3} can not reach $\Theta_2$ of Figure~\ref{sn-ex2} exactly by one-step. 
Nevertheless, applying the multiplicative $\eta$-expansion three times to $\Theta_3$, 
we can obtain $\Theta_4$ and applying the multiplicative rewrite rule four times
and the ID rewrite rule on atomic formulas five times to $\Theta_4$ of 
Figure~\ref{sn-ex4}, 
we can obtain $\Theta_2$. \\
We also give another example. 
Figure~\ref{sn-ex1-dash}, Figure~\ref{sn-ex2-dash}, and Figure~\ref{sn-ex3-dash}
also show a counterexample of one-step confluence in the IMLL system 
with the multiplicative $\eta$-expansion, since $\Theta'_3$ of Figure~\ref{sn-ex3-dash} can not reach $\Theta'_2$ of Figure~\ref{sn-ex2-dash} exactly by one-step. 
Although we can obtain $\Theta'_2$ from $\Theta'_3$ 
by applying the multiplicative rewrite rule two times and 
the ID rewrite rule two times, 
we can also obtain $\Theta'_2$ from $\Theta'_3$, 
first obtaining $\Theta'_4$ of Figure~\ref{sn-ex4-dash} from $\Theta'_3$ 
by the multiplicative $\eta$-expansion three times and 
second applying the multiplicative rule six times and the ID rule ten times.
\\
In the following we formalize the intuition.

\begin{definition}[the maximal $\eta$-expansion of an ID-link]
Let $\Theta$ be the IMLL proof net consisting of exactly one ID-link with $A^+$ and $A^-$ as the conclusions.
The maximal $\eta$-expansion of $\Theta$ is the IMLL proof net exactly with $A^+$ and
$A^-$ as the conclusions that does not have any ID-links except ID-links with
only atomic conclusions obtained from $\Theta$ by applying multiplicative $\eta$-expansion rules maximally. 
We denote the $\eta$-expansion of $\Theta$ by $\eta$-expand($A^+,A^-$).
\end{definition}

\begin{lemma}
\label{contractetaexpansion}
Let $\Pi$ be an IMLL proof net with $A^+$ (respectively $A^-$) as a conclusion.
Then we let $\Theta$ be the IMLL proof net connecting $\Pi$ and 
$\eta$-expand($A^+,A^-$) by a Cut-link with $A^+$ (respectively $A^-$) on $\Pi$
and $A^-$ (respectively $A^+$) on $\eta$-expand($A^+,A^-$) as the premises.
Then there is an IMLL proof net $\Pi'$ such that 
$\Pi \to^\ast \Pi'$ and $\Theta \to^\ast \Pi'$, where
$\Pi'$ is an IMLL proof net obtained from $\Pi$ by applying 
the multiplicative $\eta$-expansion 
to some (possibly zero) subformula occurrences of $A^+$ (resp. $A^-$) of $\Pi$.
\end{lemma}

\begin{proof}
We prove this lemma by induction on $A^+$ (resp. $A^-$). 
We only consider $A^+$. The case of $A^-$ is similar.
\begin{enumerate}
\item The base step: the case where $A^+$ is an atomic formula $p^+$.\\
Then $\eta$-expand($A^+,A^-$) is an IMLL proof net consisting exactly one ID-link with
$p^+,p^-$ as the conclusions.
Then we can easily see that $\Theta \to \Pi$ by ID rewrite rule.
So, it is OK to  let $\Pi'$ be $\Pi$.
\item The induction step: the case where $A^+$ is not an atomic formula.
\begin{enumerate}
  \item the case where $A^+$ on $\Pi$ is a conclusion of an ID-link:\\
    Let $\Pi'$ be the IMLL proof net obtained from $\Pi$ by 
    replacing the ID-link with $\eta$-expand($A^+,A^-$).
    Then $\Pi \to^\ast \Pi'$.
    Moreover it is easily see to $\Theta \to \Pi'$ by the ID rewrite rule.
  \item the case where $A^+$ on $\Pi$ is not a conclusion of an ID-link:
    \begin {enumerate}
      \item the case where $A^+$ is a conclusion of $\PAR$-link:\\
	Then $A^+$ must have the form ${A_1 \LIMP  A_2}^+$. 
	Let $\Theta'$ be the IMLL proof net such that
	$\Theta \to \Theta'$ by the multiplicative rewrite rule 1.
	Then the graph $\Theta''$ obtained from $\Theta'$ by removing
	$\PAR$-link with the conclusion ${A_1 \LIMP  A_2}^+$ is a 
	subproof net of $\Theta'$. 
	Then $\Theta''$ can be regarded as an IMLL proof net
	obtained from an IMLL proof net and $\eta$-expand($A_1^+,A_1^-$) 
	by connecting a Cut-link. 
	Let $\Pi_1$ be the IMLL proof net obtained from $\Theta''$ 
	by removing $\eta$-expand($A_1^+,A_1^-$) and its associated Cut-link.
	By inductive hypothesis, we can obtain an IMLL proof net $\Pi'_1$ 
	such that $\Pi_1 \to^\ast \Pi'_1$ and $\Theta' \to^\ast \Pi'_1$, where
	$\Pi'_1$ is obtained from $\Pi_1$ by applying 
	the multiplicative $\eta$-expansion  
	to some subformula occurrences of $A_1^-$ of $\Pi_1$.
	Again $\Pi'_1$ can be regarded as an IMLL proof net
	obtained from an IMLL proof net and $\eta$-expand($A_2^+,A_2^-$) 
	by connecting a Cut-link.
	Let $\Pi_2$ be the IMLL proof net obtained from $\Pi'_1$ 
	by removing $\eta$-expand($A_2^+,A_2^-$)  and its associated Cut-link.
	By inductive hypothesis again, we can obtain an IMLL proof net $\Pi'_2$ 
	such that $\Pi_2 \to^\ast \Pi'_2$ and $\Pi'_1 \to^\ast \Pi'_2$, where
	$\Pi'_2$ is obtained from $\Pi_2$ by applying 
	the multiplicative $\eta$-expansion 
	to some subformula occurrences of $A_2^+$ of $\Pi_1$.
	Finally let the IMLL proof net obtained from $\Pi'_2$ by adding 
	$\PAR$-link with the conclusion ${A_1 \LIMP  A_2}^+$ be $\Pi'$.
        It can be easily seen that $\Theta \to^\ast \Pi'$, $\Pi \to^\ast \Pi'$, 
	and $\Pi'$ is obtained from 
	$\Pi$  by applying 
	the multiplicative $\eta$-expansion 
	to some subformula occurrences of ${A_1 \LIMP  A_2}^+$ of $\Pi$.
      \item the case where $A^+$ is a conclusion of $\TENS$-link:\\
	Then $A^+$ must have the form ${A_1 \TENS  A_2}^+$. 
	Let $\Theta'$ be the IMLL proof net such that
	$\Theta \to \Theta'$ by the multiplicative rewrite rule 2.
	On the other hand there is an IMLL subproof net $\Pi_1$ (resp. $\Pi_2$) of $\Pi$ (and also of $\Theta'$) such that 
	$\Pi_1$ (resp. $\Pi_2$) is the maximal subproof net of $\Pi$ among the subproof nets with with a conclusion $A_1^+$ (resp. $A_2^+$)\footnote{Such a maximal subproof net is called ``empire'' in the literature (see \cite{Gir87})}. 
	Let the IMLL proof net 
	obtained by connecting $\Pi_1$ (resp. $\Pi_2$) 
	and $\eta$-expand($A_1^+,A_1^-$) (resp. $\eta$-expand($A_2^+,A_2^-$)) by
	a Cut-link be $\Theta_1$ (resp. $\Theta_2$).
	$\Theta_1$ and $\Theta_2$ is also an IMLL subproof net of $\Theta'$.
	By applying inductive hypothesis to $\Theta_1$ (resp. $\Theta_2$)
	and $\Pi_1$ (resp. $\Pi_2$),
	we obtain $\Pi'_1$ (resp. $\Pi'_2$) from $\Pi_1$ (resp. $\Pi_2$) by 
	some $\eta$-expansions such that
	$\Pi_1 \to^\ast \Pi'_1$ (resp. $\Pi_2 \to^\ast \Pi'_2$) 
        and $\Theta_1 \to^\ast \Pi'_1$ (resp. $\Theta_2 \to^\ast \Pi'_2$).
	The IMLL proof net obtained from $\Theta'$ by replacing 
	$\Theta_1$ and $\Theta_2$ by $\Pi'_1$ and $\Pi'_2$ is 
	an IMLL proof net obtained from $\Pi$ 
	by applying 
	the multiplicative $\eta$-expansion 
	to some subformula occurrences of ${A_1 \TENS A_2}^+$ of $\Pi$.
    \end{enumerate}
  \end{enumerate}
\end{enumerate}
$\Box$
\end{proof}

\begin{lemma}[Weak Confluence]
\label{weakconfluence}
In the IMLL system we assume that $\Theta \to \Theta_1$ and $\Theta \to \Theta_2$. Then there is an IMLL proof net $\Theta_3$ such that
$\Theta_1 \to^\ast \Theta_3$ and $\Theta_2 \to^\ast \Theta_3$.
\end{lemma}

\begin{figure}[htbp]
\begin{center}
\includegraphics[scale=.5]{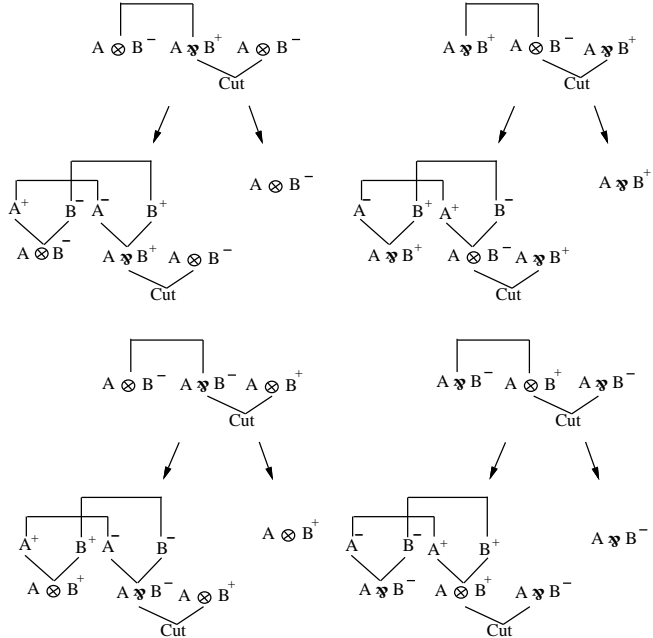}
\caption[all the critical pairs]
{all the critical pairs}
\label{all-critical}
\end{center}
\end{figure}

\begin{proof}
The problematic cases are four critical pairs in Figure~\ref{all-critical}.
Let $\Theta_1$ be the left contractum in the pairs and $\Theta_2$ be 
the right contractum. 
Then we let $\Theta'_1$ be the IMLL proof net obtained from $\Theta_1$
by applying
the multiplicative $\eta$-expansion to $\Theta_1$ 
until there are no any ID-links with non-atomic conclusions. 
Note that $\Theta_1 \to^\ast \Theta_1$.
Next we apply Lemma~\ref{contractetaexpansion} to $\Theta'_1$. 
Then we can find $\Theta_3$ such that  $\Theta_2 \to^\ast \Theta_3$. Hence $\Theta'_1 \to^\ast \Theta_3$.
$\Box$
\end{proof}

\begin{proposition}[Confluence]
The IMLL system is confluent.
\end{proposition}
\begin{proof}
From Proposition~\ref{SNIMLL} and Lemma~\ref{weakconfluence} by Newman's Lemma.
$\Box$
\end{proof}

\section{An equality on closed IMLL proof nets} \label{EQUALITY}
In this section, we define an equality on closed IMLL proof nets. 

\begin{definition}
An IMLL proof net $\Theta$ is closed if $\Theta$ has exactly one conclusion.
\end{definition}
Next we consider the forms of normal IMLL proof nets.
Let $\Theta$ be a normal IMLL proof net with the positive conclusion $A^+$ and 
the other conclusions $B_1^-, \cdots, B_n^-$.

\begin{figure}[htbp]
\begin{center}
\includegraphics[scale=.5]{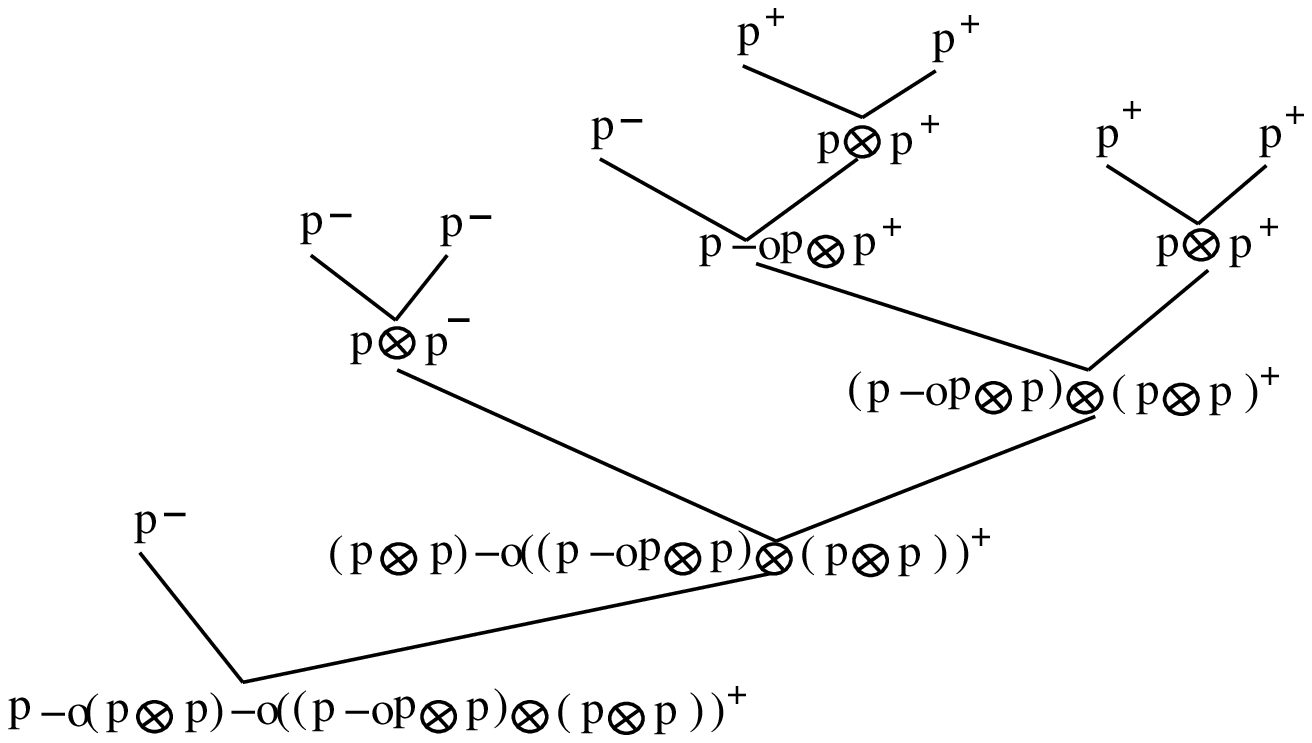}
\caption[the abstract syntax tree of ${p \LIMP (p \TENS p) \LIMP ((p \LIMP p \TENS p) \TENS (p \TENS p))}^+$]
{the abstract syntax tree of ${p \LIMP (p \TENS p) \LIMP ((p \LIMP p \TENS p) \TENS (p \TENS p))}^+$}  
\label{imll-abstract-tree-ex1}
\end{center}
\end{figure}

\begin{figure}[htbp]
\begin{center}
\includegraphics[scale=.5]{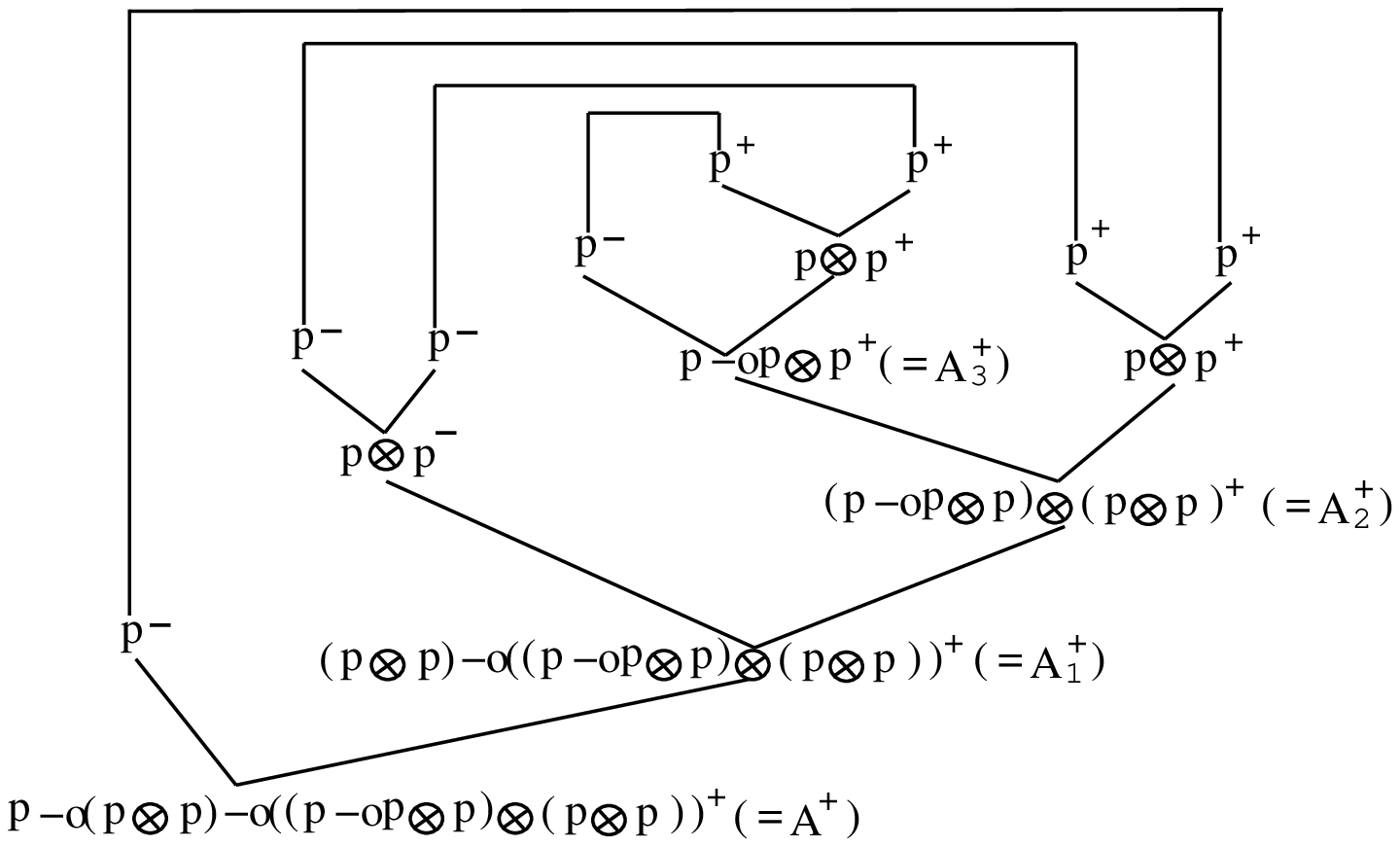}
\caption[a closed IMLL proof net of ${p \LIMP (p \TENS p) \LIMP ((p \LIMP p \TENS p) \TENS (p \TENS p))}^+$]
{a closed IMLL proof net of  ${p \LIMP (p \TENS p) \LIMP ((p \LIMP p \TENS p) \TENS (p \TENS p))}^+$}  
\label{imll-pn-ex1}
\end{center}
\end{figure}

We consider the unique abstract syntax forest  $T(A^+), T(B_1^-), \cdots, T(B_n^-)$ determined by 
$A^+,B_1^-, \cdots, B_n^-$, where $T(A^+)$ (resp. $T(B_i^-)$ $\, (1 \le i \le n)$) is 
the unique abstract syntax tree determined by $A^+$ (resp. $B_i^-$ $\, (1 \le i \le n)$).
For example, when let $A^+$ be
${p \LIMP (p \TENS p) \LIMP ((p \LIMP p \TENS p) \TENS (p \TENS p))}^+$, 
Figure~\ref{imll-abstract-tree-ex1} is  the abstract syntax tree $T(A^+)$.\\
Then we define a set ${\bf P}_{\Theta}$ of alternating sequences 
of nodes of the forest  $T(A^+), T(B_1^-), \cdots, T(B_n^-)$ and $\{ {\bf L}, {\bf R}, {\bf ID} \}$ 
as follows:
\begin{enumerate}
\item $A^+ \in {\bf P}_{\Theta}$;
\item If $s, {A_1 \TENS A_2}^+ \in {\bf P}_{\Theta}$, where $s$ is an alternating sequence, then
  $s, {A_1 \TENS A_2}^+, {\bf L}, A_1^+ \in {\bf P}_{\Theta}$ and $s, {A_1 \TENS A_2}^+, {\bf R}, A_2^+ \in {\bf P}_{\Theta}$;
\item If $s, {A_1 \LIMP A_2}^+ \in {\bf P}_{\Theta}$, then  $s, {A_1 \LIMP A_2}^+, {\bf R}, A_2^+ \in {\bf P}_{\Theta}$;
\item If $s, p^+ \in  {\bf P}_{\Theta}$, then $s, p^+, {\bf ID}, p^- \in  {\bf P}_{\Theta}$;
\item If $s, {A'}^- \in {\bf P}_{\Theta}$ and ${A'}^-$ is the right premise of a $\TENS^-$-link $L$, then
  $s, {A'}^-, {\bf R}, {{A''} \TENS {A'}}^- \in {\bf P}_{\Theta}$, where 
  ${{A''} \TENS {A'}}^-$ is the conclusion of $L$;
\item If $s, {A'}^- \in {\bf P}_{\Theta}$ and ${A'}^-$ is the left premise of a $\PAR^-$-link $L$, then
  $s, {A'}^-, {\bf L}, {{A'} \PAR {A''}}^- \in {\bf P}_{\Theta}$, where 
  ${{A'} \PAR {A''}}^-$ is the conclusion of $L$;
\item If $s, {A'}^- \in {\bf P}_{\Theta}$ and ${A'}^-$ is the right premise of a $\PAR^-$-link $L$, then
  $s, {A'}^-, {\bf R}, {{A''} \PAR {A'}}^- \in {\bf P}_{\Theta}$, where 
  ${{A''} \PAR {A'}}^-$ is the conclusion of $L$.
\end{enumerate}
We say that $s, B^- \in {\bf P}_{\Theta}$ is a {\it main path} of $\Theta$, if 
$B^-$ is neither a premise of $\TENS^-$-link nor $\PAR^-$-link in $\Theta$.
Then we call $B^-$ {\it the head} of the main path.
Note that if $\Theta$ is an IIMLL proof net, then $\Theta$ has exactly one main path.
If the positive conclusion of a subproof net of $\Theta$ is 
the left premise of a $\TENS^-$-link in a main path, 
then we call the subproof net {\it a direct subproof net} of $\Theta$.\\
For example, Figure~\ref{imll-pn-ex1} shows a closed IMLL proof net 
of ${p \LIMP (p \TENS p) \LIMP ((p \LIMP p \TENS p) \TENS (p \TENS p))}^+$, 
where we give abbreviations to some formula occurrences. 
There are exactly four main paths in the IMLL proof net:
\begin{enumerate}
\item \label{mainpath1} $A^+, {\bf R}, A_1^+, {\bf R}, A_2^+, {\bf L}, A_3^+, {\bf R}, {p \TENS p}^+, {\bf L}, p^+, {\bf ID}, p^-$
\item $A^+, {\bf R}, A_1^+, {\bf R}, A_2^+, {\bf L}, A_3^+, {\bf R}, {p \TENS p}^+, {\bf R}, p^+, {\bf ID}, p^-, {\bf R}, {p \TENS p}^-$
\item \label{mainpath3} $A^+, {\bf R}, A_1^+, {\bf R}, A_2^+, {\bf R}, {p \TENS p}^+, {\bf L}, p^+, {\bf ID}, p^-, {\bf L}, {p \TENS p}^-$
\item $A^+, {\bf R}, A_1^+, {\bf R}, A_2^+, {\bf R}, {p \TENS p}^+, {\bf R}, p^+, {\bf ID}, p^-$
\end{enumerate}
The head of the path (\ref{mainpath3}) is ${p \TENS p}^-$.
Note that there is no direct subproof net of the IMLL proof net.

\begin{figure}[htbp]
\begin{center}
\includegraphics[scale=.5]{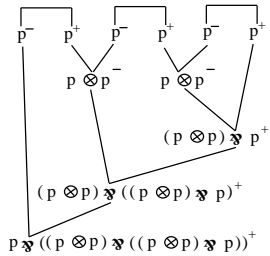}
\caption[an IMLL proof net]{an IMLL proof net}  
\label{eq-justify}
\end{center}
\end{figure}

Next, we define an equality on normal IMLL proof nets. 
Since we define IMLL proof nets inductively,
it seems a reasonable definition that
two proof nets are equal, if 
these are the same w.r.t forms and orders of applied rules in Figure~\ref{imllpn}.
But if we defined an equality in this way, 
then there would be two different IMLL proof nets with 
the form of Figure~\ref{eq-justify}, 
since there are two orders of applied rules in order to 
define the IMLL proof net. 
Because this is unreasonable, we define an equality in the following way.

\begin{figure}[htbp]
\begin{center}
\includegraphics[scale=.5]{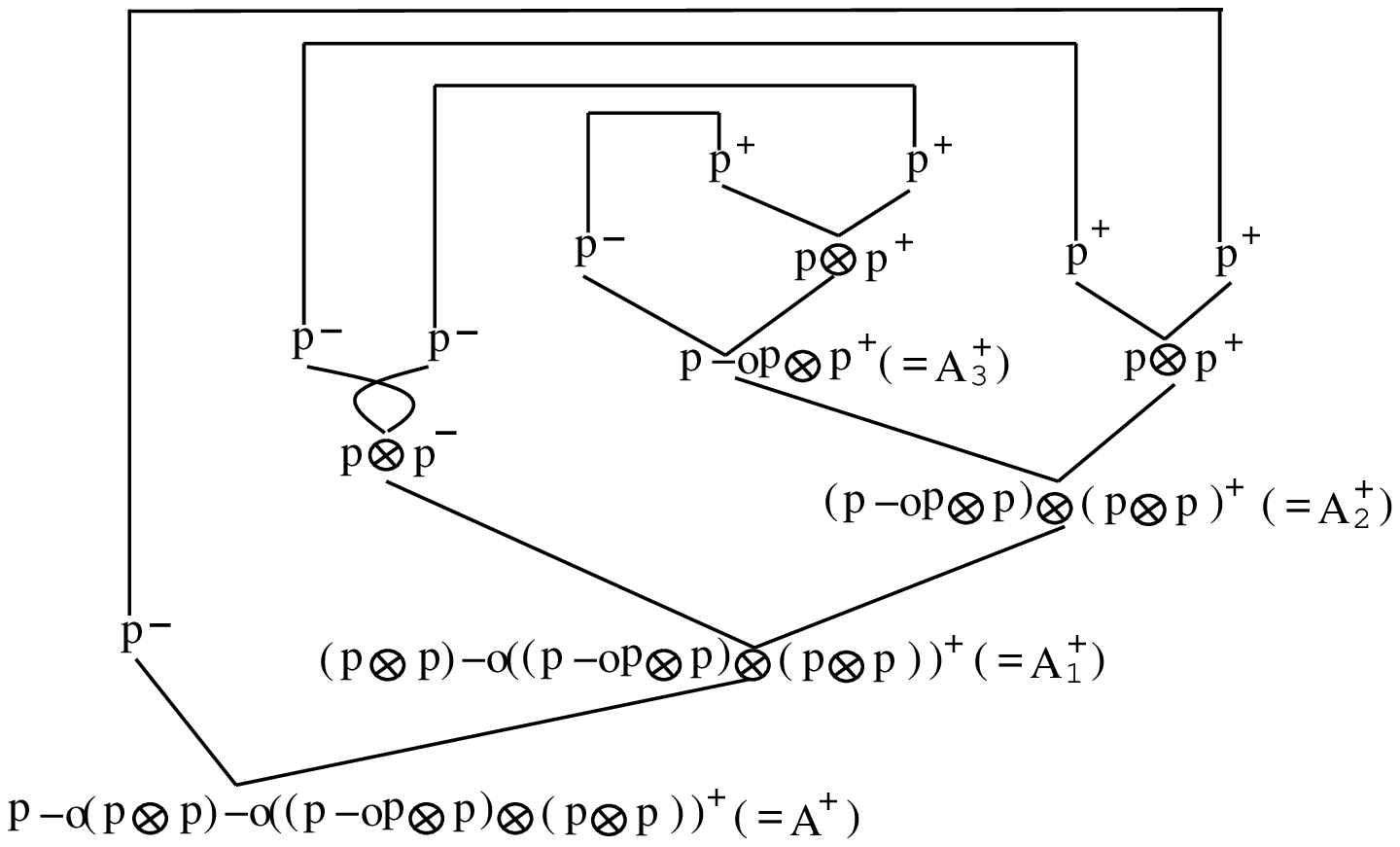}
\caption[another closed IMLL proof net of ${p \LIMP (p \TENS p) \LIMP ((p \LIMP p \TENS p) \TENS (p \TENS p))}^+$]
{another closed IMLL proof net of  ${p \LIMP (p \TENS p) \LIMP ((p \LIMP p \TENS p) \TENS (p \TENS p))}^+$}  
\label{imll-pn-ex2}
\end{center}
\end{figure}

\begin{definition}[an equality on normal IMLL proof nets]
Let $\Theta_1$ and $\Theta_2$ be two normal IMLL proof nets with the same 
positive conclusion.
Then $\Theta_1 = \Theta_2$ if 
\begin{enumerate}
\item For each main path of $\Theta_1$ there is completely the same main path in $\Theta_2$.
Moreover there is no any path in $\Theta_2$ other than these corresponding paths, 
i.e., there is a bijection from the set of the main paths of $\Theta_1$ to that of $\Theta_2$, 
which can be regarded as an identity map and
\item The head of a main path in $\Theta_1$ is a premise of a $\PAR^+$-link $L'$ iff 
the corresponding head of  $\Theta_2$ is also a premise of the $\PAR^+$-link $L''$ with the same position as $L'$ and
\item If a direct subproof net of $\Theta_1$ is $\Theta'$ and 
the corresponding subproof net of $\Theta_2$ is $\Theta''$, then
$\Theta' = \Theta''$ and
\item A head of a subproof net of $\Theta_1$ is 
a premise of a $\PAR^+$-link $L'$ in a main path of $\Theta_1$ iff 
that of the corresponding subproof net of $\Theta_2$ is also
a premise of the $\PAR^+$-link $L''$ with the same position as $L'$.
\end{enumerate}
\end{definition}
For example, the IIMLL proof net of Figure~\ref{imll-pn-ex1} (let the net be $\Theta_1$) and that of Figure~\ref{imll-pn-ex2} (let the net be $\Theta_2$) 
are two IMLL proof nets with the same conclusion. 
But $\Theta_1 \neq \Theta_2$, because 
there is no corresponding path in $\Theta_2$ to the path 
$A^+, {\bf R}, A_1^+, {\bf R}, A_2^+, {\bf R}, {p \TENS p}^+, {\bf L}, p^+, {\bf ID}, p^-, {\bf L}, {p \TENS p}^-$
in $\Theta_1$. \\
If the structure of proof nets is forgotten and collapses to the usual lambda calculus (see \cite{Gir98}), our equality corresponds to 
the union of the usual $\beta\eta$-equality and the equivalence up to  
bijective replacement of free variables. 
But also note that our equality is not that of proof nets as graphs: 
for example, if we consider graphs whose nodes are links and whose edges
are formulas (i.e., Danos-Regnier style's proof-nets, see \cite{DR95}), those of Figure~\ref{imll-pn-ex1} and Figure~\ref{imll-pn-ex2} are equal, 
because such graphs have no information about whether a premise of a link is left or right. 
On the other hand, it has a subtle point to extend our equality to the fragment including 
the multiplicative constant ${\bf 1}$: the topic will be given elsewhere.
\section{Third-order reduction on IIMLL proof nets} \label{THIRDRED}
\begin{figure}[htbp]
\begin{center}
\includegraphics[scale=.5]{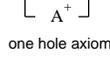}
\caption[one-hole axiom link]{one-hole axiom link}
\label{holeaxiom}
\end{center}
\end{figure}

In this section and the next section  we only consider IIMLL proof nets. 
We assume that we are given two closed IIMLL proof nets $\Theta_1$ and $\Theta_2$ with the same conclusion such that $\Theta_1 \neq \Theta_2$.
In this section we show that we can find a context $C[]$ such that $C[\Theta_1]$ and $C[\Theta_2]$ have different normal forms and orders less than 4-th order.
\begin{definition}[hole axioms]
A hole axiom with the positive conclusion $A^+$ is a link 
with the form shown by Figure~\ref{holeaxiom}.
\end{definition}

\begin{figure}[htbp]
\begin{center}
\includegraphics[scale=.5]{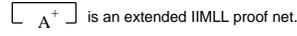}
\caption[extended IIMLL proof nets]{extended IIMLL proof nets}
\label{exiimllpn}
\end{center}
\end{figure}

\begin{definition}[extended IIMLL proof nets and one-hole contexts]
Extended IIMLL proof nets are inductively defined 
by using the rules of Figure~\ref{imllpn} except for clauses (4) and (6) and that of Figure~\ref{exiimllpn}.
A one-hole context (for short context) is an extended IIMLL proof net with
exactly one one-hole axiom.
\end{definition}
We use  $C[], C_0[], C_1[], \ldots$ to denote one-hole contexts.

\begin{remark}
Unlike \cite{Bar84}, there is no capture of free variables with regard to
our notion of contexts,
since we are working on closed proof nets.
\end{remark}
\begin{definition}
Let $\Theta$ be an IIMLL proof net with the positive conclusion $A^+$
and $C[]$ be a one-hole context with the one-hole axiom $A^+$. 
Then $C[\Theta]$ is an IIMLL proof net obtained from $C[]$ by 
replacing one-hole axiom $A^+$ by $\Theta$.
\end{definition}

\begin{definition}[depth]
The depth of an IIMLL proof net $\Theta$ (denoted by $\depth(\Theta)$) is inductively defined as follows:
\begin{enumerate}
\item If the main path of $\Theta$ does not include $\TENS^-$-links, then 
$\depth(\Theta)$ is $1$.
\item Otherwise, 
when all the direct subproof nets of $\Theta$ are 
$\Theta_1, \ldots, \Theta_m$,
$\depth(\Theta)$ is $\max \{\depth(\Theta_1), \ldots, \depth(\Theta_m) \} + 1$.
\end{enumerate}
The depth of a positive formula occurrence $A^+$ in $\Theta$ is $\depth(\Theta) - \depth(\Theta') + 1$, where 
$\Theta'$ is the subproof net of $\Theta$ which is the least among subproof nets including $A^+$.
\end{definition}

\begin{definition}[the order of a positive IIMLL formula]
The order of an IIMLL formula $A^+$, denoted by $\order(A^+)$ is inductively as follows:
\begin{enumerate}
\item If $A^+$ is an atomic formula $p^+$  then 
$\order(A^+)$ is $1$.
\item If $A^+$ is ${A_1 \LIMP \ldots \LIMP A_n \LIMP p}^+$, then 
$\order(A^+)$ is 
\[ \max \{ \order(A_1^+), \ldots, \order(A_n^+) \}+1. \]
\end{enumerate}
\end{definition}
We define the {\it order} of a closed IIMLL proof net $\Theta$ as the order of the positive conclusion.
\begin{definition}[the measure w.r.t linear implication]
Let $\Theta$ be an IIMLL proof net.
The measure of $\Theta$ w.r.t linear implication denoted by $\measure_{\LIMP}(\Theta)$ is
the sum of depths of all the positive formula occurrences of $\Theta$.
\end{definition}

\begin{figure}[htbp]
\begin{center}
\includegraphics[scale=.5]{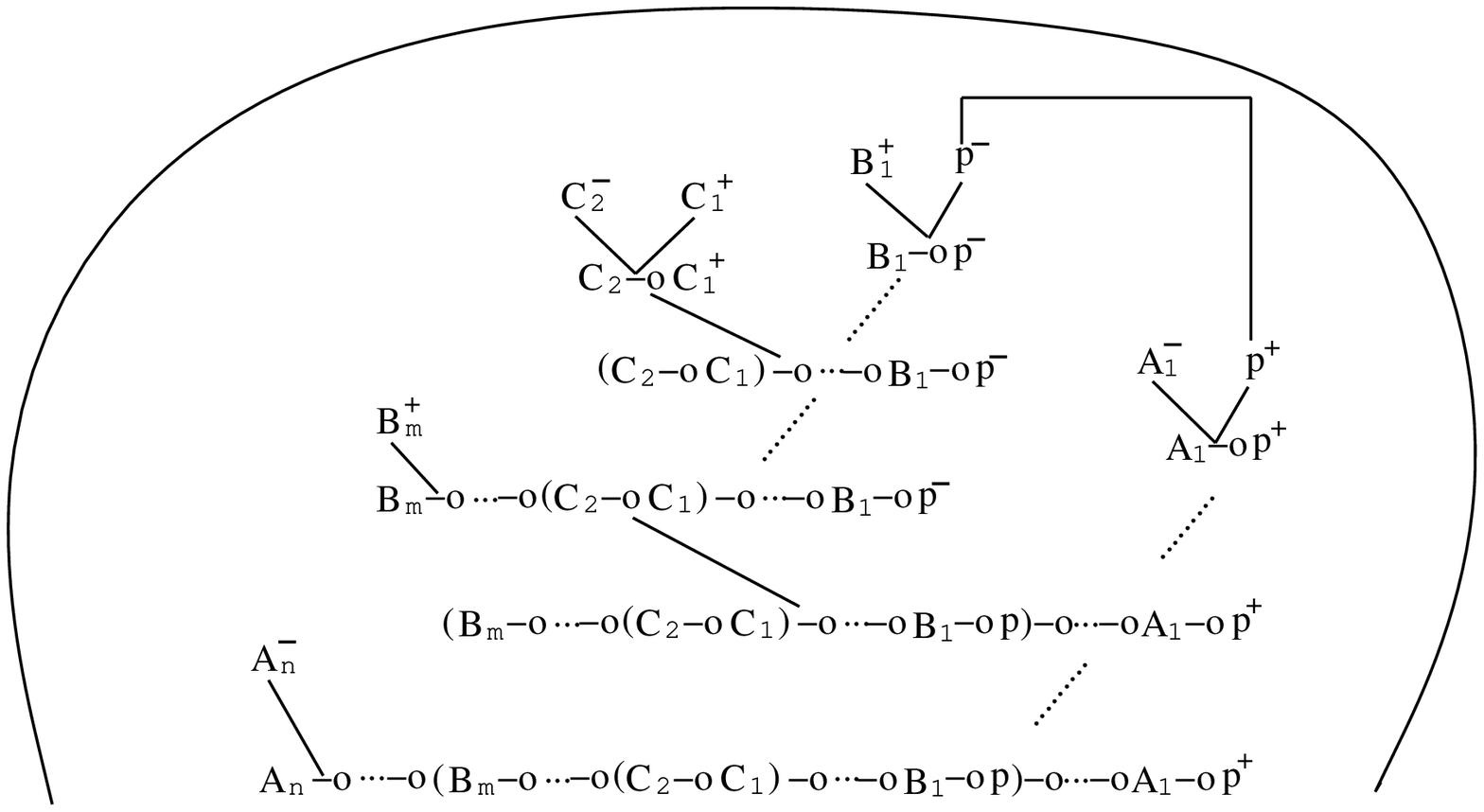}
\caption[An IIMLL proof net before reduced]{An IIMLL proof net before reduced}
\label{reductionlemmaassump}
\end{center}
\end{figure}

\begin{lemma}
\label{reductionlemma}
Let $\Theta$ be an IIMLL proof net with the positive conclusion 
\[ {A_n \LIMP \ldots \LIMP
(B_m \LIMP \ldots \LIMP (C_2 \LIMP C_1) \LIMP \ldots \LIMP B_1 \LIMP p)
\LIMP \ldots \LIMP A_1 \LIMP p}^+ \]
and the form shown in Figure~\ref{reductionlemmaassump}.
Then there is an IIMLL proof net with the positive conclusion
\[ {C_2 \LIMP A_n \LIMP \ldots \LIMP
(B_m \LIMP \ldots \LIMP C_1 \LIMP \ldots \LIMP B_1 \LIMP p)
\LIMP \ldots \LIMP A_1 \LIMP p}^+. \]
\end{lemma}

\begin{figure}[htbp]
\begin{center}
\includegraphics[scale=.5]{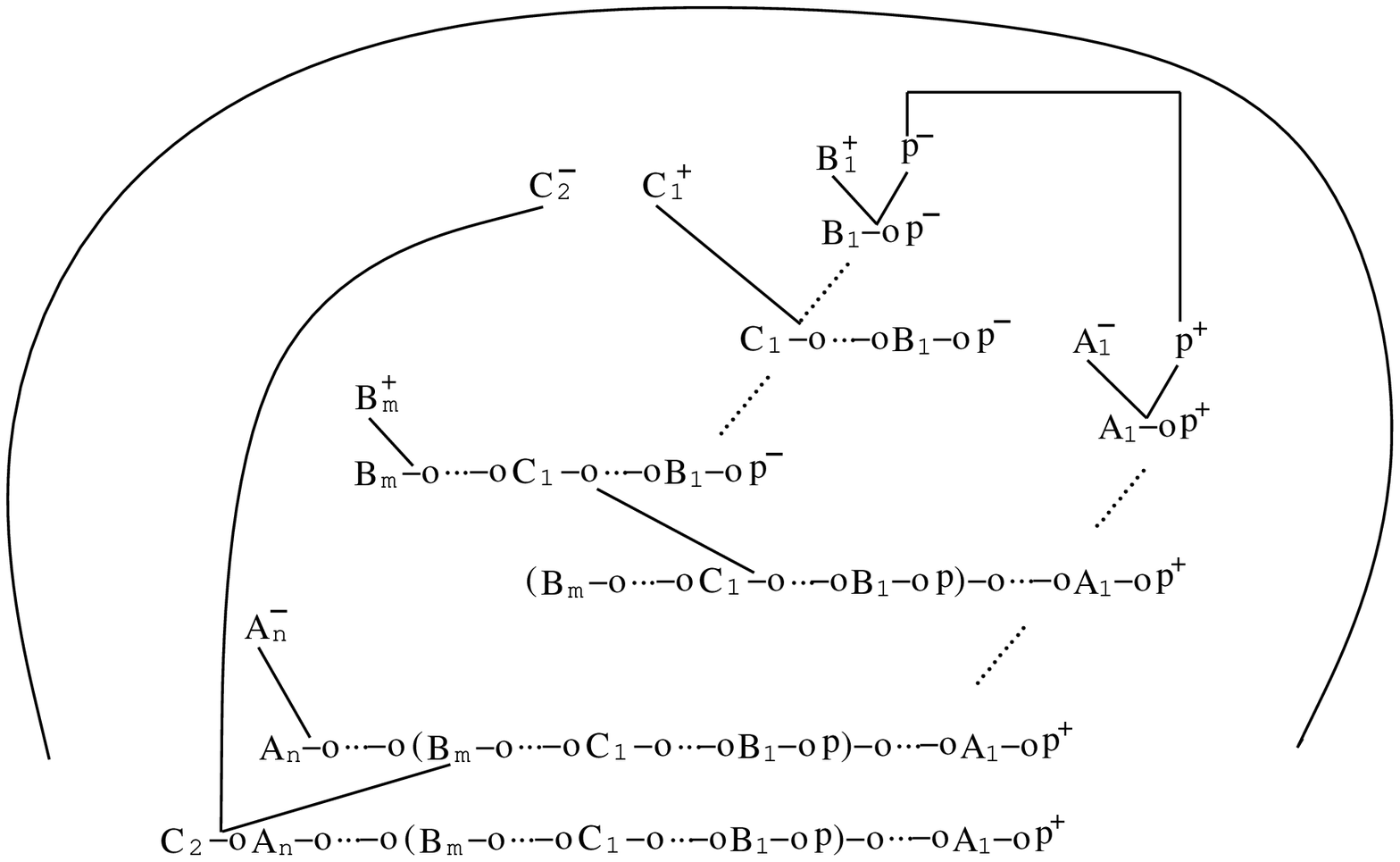}
\caption[The IIMLL proof net after reduced]{The IIMLL proof net after reduced}
\label{reductionlemmaconc}
\end{center}
\end{figure}

\begin{proof}
The proof structure of Figure~\ref{reductionlemmaconc} obtained from Figure~\ref{reductionlemmaassump}
by manipulating some links is also an IIMLL proof net (the invisible part of $\Theta$ is never touched), 
because all the Danos-Regnier graphs of the IMLL proof structure of Figure~\ref{reductionlemmaconc} can be regarded as 
a subset of that of Figure~\ref{reductionlemmaassump} in the following way:
\begin{itemize}
	\item In the $\PAR$-link with the conclusion 
${C_2 \LIMP A_n \LIMP \ldots \LIMP
(B_m \LIMP \ldots \LIMP C_1 \LIMP \ldots \LIMP B_1 \LIMP p)
\LIMP \ldots \LIMP A_1 \LIMP p}^+$, 
if $C_2^-$ is chosen, then identify $C_2^-$ with the conclusion of the $\PAR$-link;
\item otherwise, identify the other premise with the conclusion of the $\PAR$-link. 
\end{itemize}
If the proof structure of Figure~\ref{reductionlemmaconc} were not 
an IMLL proof net, that is, did not satisfy the criterion of Theorem~\ref{seqthm}, then 
$\Theta$ would not be an IMLL proof net by Theorem~\ref{seqthm}. This is a contradiction.
$\Box$
\end{proof}

\begin{proposition}
\label{redprop}
Let $\Theta_1$ and $\Theta_2$ be two closed IIMLL proof nets with the same positive conclusion and 
an order greater than 3 such that $\Theta_1 \neq \Theta_2$.
Then there is a context $C[]$ such that 
$\measure_{\LIMP}(\Theta_1) > \measure_{\LIMP}(C[\Theta_1])$,
$\measure_{\LIMP}(\Theta_2) > \measure_{\LIMP}(C[\Theta_2])$,
and $C[\Theta_1] \neq C[\Theta_2]$.
\end{proposition}

\begin{proof}
Since $\Theta_1$ has an order greater than 3, 
the positive conclusion of $\Theta_1$ has the form
\[ {A_n \LIMP \ldots \LIMP
(B_m \LIMP \ldots \LIMP B_{k+1} \LIMP (C_2 \LIMP C_1) \LIMP B_{k-1} \LIMP \ldots \LIMP B_1 \LIMP p)
\LIMP \ldots \LIMP A_1 \LIMP p}^+ \]
for some $k$ ($1 \le k \le m$).

On the other hand, there is an IIMLL proof net that is $\eta$-expansion of ID-link with the conclusion
\[ {A_n \LIMP \ldots \LIMP
(B_m \LIMP \ldots \LIMP B_{k+1} \LIMP (C_2 \LIMP C_1) \LIMP B_{k-1} \LIMP \ldots \LIMP B_1 \LIMP p)
\LIMP \ldots \LIMP A_1 \LIMP p}^- \]
and
\[ {A_n \LIMP \ldots \LIMP
(B_m \LIMP \ldots \LIMP B_{k+1} \LIMP (C_2 \LIMP C_1) \LIMP B_{k-1} \LIMP \ldots \LIMP B_1 \LIMP p)
\LIMP \ldots \LIMP A_1 \LIMP p}^+. \]
Then by Lemma~\ref{reductionlemma} we can obtain an IIMLL proof net $\Pi$ whose conclusions are exactly
\[ {A_n \LIMP \ldots \LIMP
(B_m \LIMP \ldots \LIMP B_{k+1} \LIMP (C_2 \LIMP C_1) \LIMP B_{k-1} \LIMP \ldots \LIMP B_1 \LIMP p)
\LIMP \ldots \LIMP A_1 \LIMP p}^- \]
and
\[ {C_2 \LIMP A_n \LIMP \ldots \LIMP
(B_m \LIMP \ldots \LIMP B_{k+1} \LIMP C_1 \LIMP B_{k-1} \LIMP \ldots \LIMP B_1 \LIMP p)
\LIMP \ldots \LIMP A_1 \LIMP p}^+. \]
Then let $C[]$ be the context obtained from $\Pi$ by connecting $\Pi$'s negative conclusion 
and one-hole axiom via Cut-link. 
Then the number of the positive formula occurrences of $\Theta_1$ is equal to that of $C[\Theta_1]$.
The positive formula occurrence ${C_1 \LIMP C_2}^+$ occurs in $\Theta_1$ and depth $2$, 
but not in $C[\Theta_1]$, 
while 
the positive formula occurrence 
\[ {C_2 \LIMP A_n \LIMP \ldots \LIMP
(B_m \LIMP \ldots \LIMP B_{k+1} \LIMP C_1 \LIMP B_{k-1} \LIMP \ldots \LIMP B_1 \LIMP p)
\LIMP \ldots \LIMP A_1 \LIMP p}^+ \]
occurs in $C[\Theta_1]$ and has depth $1$, but not in $\Theta_1$. 
The other formula occurrences in $\Theta_1$ are the same as that of $C[\Theta_1]$. 
So, it is obvious that $\measure_{\LIMP}(\Theta_1) > \measure_{\LIMP}(C[\Theta_1]).$ \\
Note that in Proposition~\ref{redprop} the construction of $C[]$ only depends on 
the selection of a positive subformula occurrence of $\Theta_1$. 
Since $\Theta_2$ has a closed IIMLL proof net with the same positive conclusion of $\Theta_1$, 
we can easily see that $\measure_{\LIMP}(\Theta_2) > \measure_{\LIMP}(C[\Theta_2])$.\\
Next in order to prove $C[\Theta_1] \neq C[\Theta_2]$, we consider the following cases:
\begin{enumerate}
\item the case where the $\PAR^+$-link of $\Theta_1$ and that of $\Theta_2$ to be manipulated by $C[]$ does not contribute to 
the unequality of $\Theta_1$ and $\Theta_2$:\\
It is obvious $C[\Theta_1] \neq C[\Theta_2]$ since 
$C[]$ does not influence the rest. 
\item the case where the $\PAR^+$-link of $\Theta_1$ and that of $\Theta_2$ to be manipulated by $C[]$ contributes to 
the unequality of $\Theta_1$ and $\Theta_2$:\\
Then, the negative premise of the $\PAR^+$-link in $\Theta_1$ differs from 
that in $\Theta_2$ (as occurrences). 
Since the position in $C[\Theta_1]$ of the manipulated $\PAR^+$-link by $C[]$ is the same as 
that in $C[\Theta_2]$ and 
the position in $C[\Theta_1]$ of the premise of the $\PAR^+$-link differs that in $C[\Theta_2]$, 
it is obvious $C[\Theta_1] \neq C[\Theta_2]$. $\Box$
\end{enumerate}
\end{proof}

\begin{figure}[htbp]
\begin{center}
\includegraphics[scale=.5]{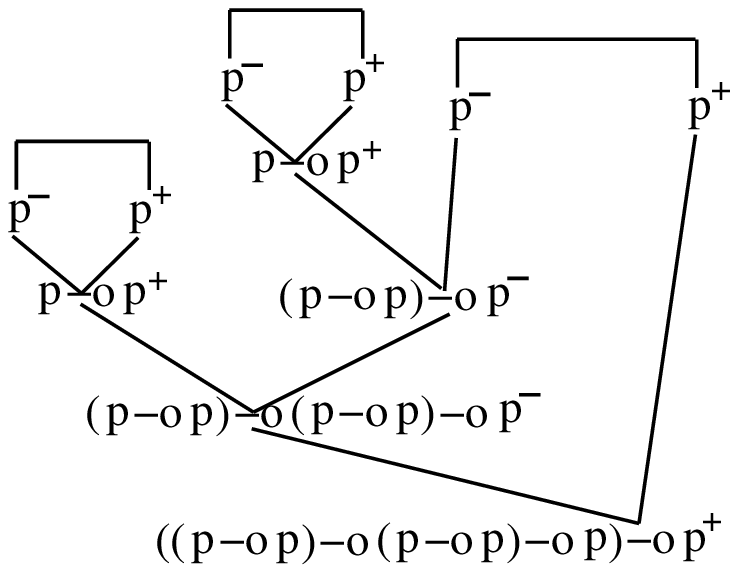}
\qquad
\qquad
\includegraphics[scale=.5]{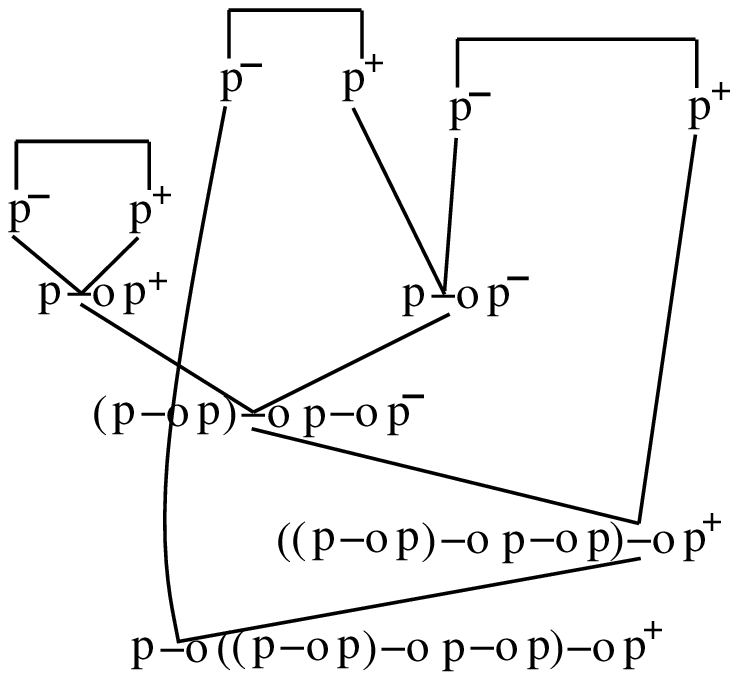}
\caption[An IIMLL proof net before reduced 
and the IIMLL proof net after reduced]
{An IIMLL proof net before reduced
and the IIMLL proof net after reduced}
\label{ex1-before-and-after}
\end{center}
\end{figure}

\begin{figure}[htbp]
\begin{center}
\includegraphics[scale=.5]{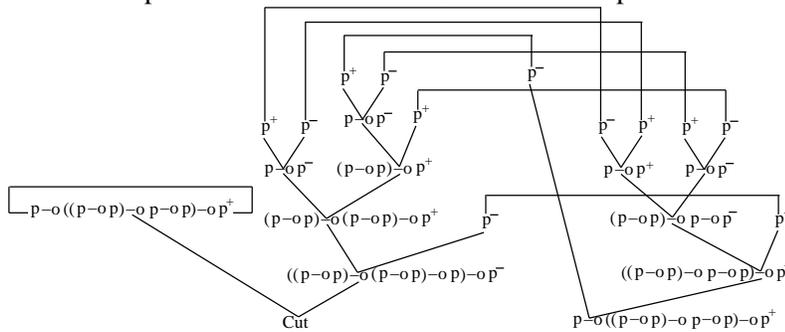}
\caption[A context]{A context}
\label{ex1-context}
\end{center}
\end{figure}

\begin{example}
Let $\Theta_1$ be the IIMLL proof net shown in the left side of Figure~\ref{ex1-before-and-after}.
Then $\measure_{\LIMP}(\Theta_1)=10$. From Proposition~\ref{redprop} we obtain 
the context shown in Figure~\ref{ex1-context}. 
By applying  the context to $\Theta_1$ and normalizing the resulting net, 
we obtain the IIMLL proof net $\Theta_2$ shown in the right side of Figure~\ref{ex1-before-and-after}.
Then $\measure_{\LIMP}(\Theta_2)=9$.
\end{example}

\begin{corollary}
\label{thirdredcontextcor}
Let $\Theta_1$ and $\Theta_2$ be closed IIMLL proof nets with the same positive conclusion and 
an order greater than 3 such that $\Theta_1 \neq \Theta_2$.
Then there is a context $C[]$ such that $C[\Theta_1] \neq C[\Theta_2]$ and both have an order less than 4.
\end{corollary}

\begin{proof}
By Proposition~\ref{redprop} we find a natural number $n$ ($n>0$) and a sequence of contexts $C_1[],C_2[],\ldots,C_n[]$ 
such that $C_1[C_2[\ldots C_n[\Theta_1] \ldots ]] \neq C_1[C_2[\ldots C_n[\Theta_2] \ldots ]]$ 
and both have an order less than 4.
Then it is obvious that there is a context $C[]$ such that $C[\Theta]=C_1[C_2[\ldots C_n[\Theta] \ldots ]]$
for any IIMLL proof net $\Theta$ with the same positive conclusion as $\Theta_1$ and $\Theta_2$. $\Box$ 
\end{proof}

\section{Value separation in third-order IIMLL proof nets} \label{VALSEP}
We assume that we are given two different normal IIMLL proof nets 
$\Theta_1$ and $\Theta_2$ with the same conclusion and with an order 
less than 4. 
However, we can not perform a separation directly. We need type instantiation.
\begin{definition}[Type instantiation]
Let $\Theta$ be an IIMLL proof net and $A$ be an MLL formula.
The type instantiated proof net $\Theta [A/p]$ of $\Theta$ w.r.t $A$ is an IIMLL proof net 
obtained from $\Theta$ by replacing each atomic formula occurrence $p$ by $A$.
\end{definition}
In the following, given two closed IIMLL proof nets $\Theta_1$ and $\Theta_2$ with the same conclusion 
and with an order less than 4 such that $\Theta_1 \neq \Theta_2$, 
we consider two type instantiated proof nets 
$\Theta_1[{\scriptstyle p \LIMP (p \LIMP p) \LIMP (p \LIMP p) \LIMP p}/p]$ 
and $\Theta_2[{\scriptstyle p \LIMP (p \LIMP p) \LIMP (p \LIMP p) \LIMP p}/p]$. 

\subsection{The definable functions on $p \LIMP (p \LIMP p) \LIMP (p \LIMP p) \LIMP p$}
\label{deffuncs}

\begin{figure}[htbp]
\begin{center}
\includegraphics[scale=.5]{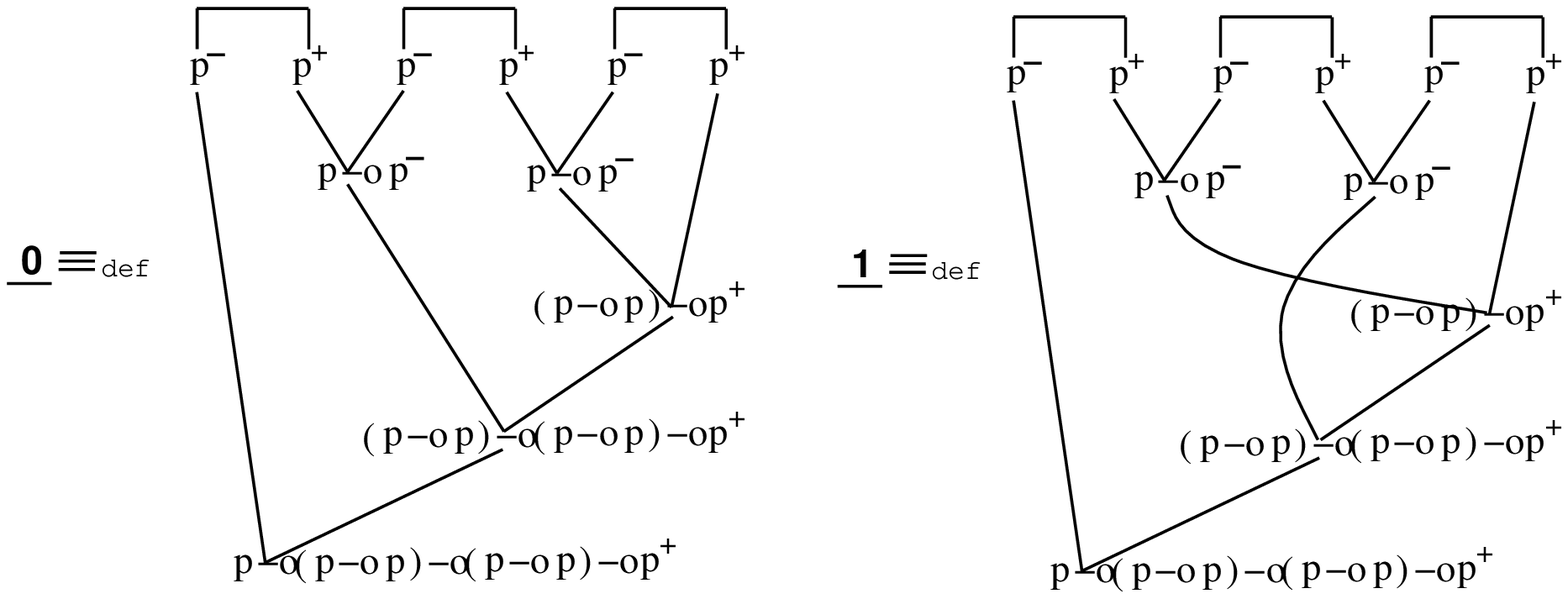}
\caption[the two normal forms on $p \LIMP (p \LIMP p) \LIMP (p \LIMP p) \LIMP p$]
{the two normal forms on $p \LIMP (p \LIMP p) \LIMP (p \LIMP p) \LIMP p$}
\label{basemultiinstant}
\end{center}
\end{figure}

Figure~\ref{basemultiinstant} shows the two closed normal proof nets on $p \LIMP (p \LIMP p) \LIMP (p \LIMP p) \LIMP p$.
We call the left proof net $\underline{0}$ and the right one $\underline{1}$.
We discuss the definable functions on $\{ \underline{0}, \underline{1} \}$ in proof nets. 

\begin{figure}[htbp]
\begin{center}
\includegraphics[scale=.5]{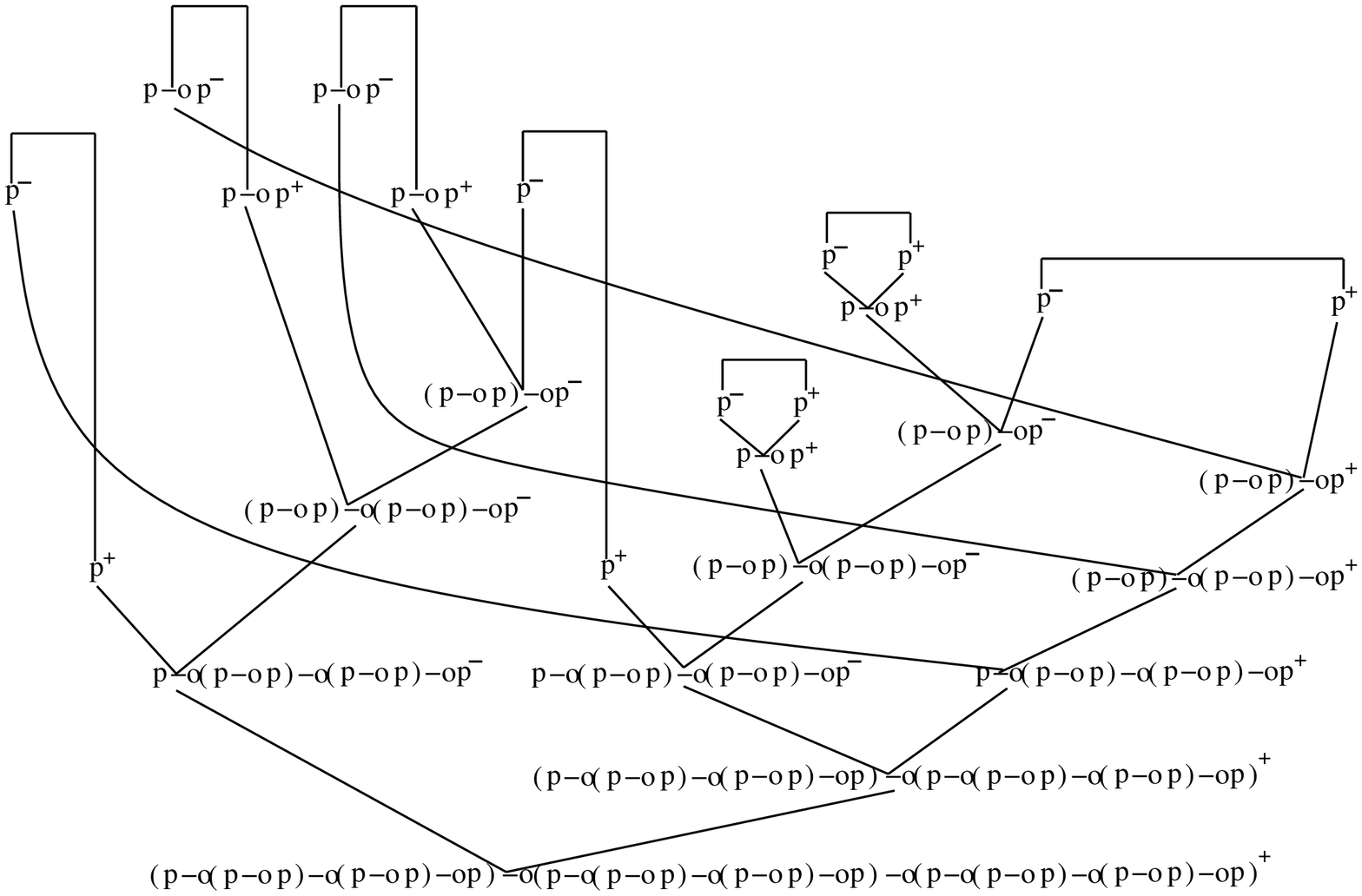}
\caption[a normal form on $\scriptstyle { (p \LIMP (p \LIMP p) \LIMP (p \LIMP p) \LIMP p) \LIMP (p \LIMP (p \LIMP p) \LIMP (p \LIMP p) \LIMP p) \LIMP (p \LIMP (p \LIMP p) \LIMP (p \LIMP p) \LIMP p)}$]
{a normal form on $\scriptstyle{ (p \LIMP (p \LIMP p) \LIMP (p \LIMP p) \LIMP p) \LIMP (p \LIMP (p \LIMP p) \LIMP (p \LIMP p) \LIMP p) \LIMP (p \LIMP (p \LIMP p) \LIMP (p \LIMP p) \LIMP p)}$}
\label{afunction-on-multiinstant}
\end{center}
\end{figure}

There are 20 closed normal proof nets of ${\scriptstyle (p \LIMP (p \LIMP p) \LIMP (p \LIMP p) \LIMP p) \LIMP (p \LIMP (p \LIMP p) \LIMP (p \LIMP p) \LIMP p)}$. 
Then we can easily see that all the one-argument functions on $\{ \underline{0}, \underline{1} \}$ are 
definable by these proof nets.\footnote{Among these 20 proof nets, 18 proof nets define a constant function 
$e_1$ or $e_2$ of Table~\ref{oneargumentfuntable}. 
A remarkable point of our separation result is 
that even if we choose two different proof nets that denote 
the same constant function among such proof nets, we can find a
context that separates these two proof nets.}
Table~\ref{oneargumentfuntable} shows these definable functions.
As to two-argument functions, 
there are 112 closed normal proof nets  
of \\
${\scriptstyle (p \LIMP (p \LIMP p) \LIMP (p \LIMP p) \LIMP p) \LIMP (p \LIMP (p \LIMP p) \LIMP (p \LIMP p) \LIMP p) \LIMP (p \LIMP (p \LIMP p) \LIMP (p \LIMP p) \LIMP p)}$.
For example, Figure~\ref{afunction-on-multiinstant} shows such a proof net.
The 112 proof nets define six two-argument functions on $\{ \underline{0}, \underline{1} \}$.
Table~\ref{twoargumentsfuntable} shows these six functions. 
In general, for any $n \, (n \ge 1)$, 
all the closed normal proof nets on \[
{\scriptstyle \overbrace{\scriptstyle (p \LIMP (p \LIMP p) \LIMP (p \LIMP p) \LIMP p) \LIMP \ldots \LIMP (p \LIMP (p \LIMP p) \LIMP (p \LIMP p) \LIMP p)}^n \LIMP (p \LIMP (p \LIMP p) \LIMP (p \LIMP p) \LIMP p)} \]
define $2n+2$ functions.\footnote{The number of the closed normal proof nets of ${\scriptstyle \overbrace{\scriptstyle (p \LIMP (p \LIMP p) \LIMP (p \LIMP p) \LIMP p) \LIMP \ldots \LIMP (p \LIMP (p \LIMP p) \LIMP (p \LIMP p) \LIMP p)}^n \LIMP (p \LIMP (p \LIMP p) \LIMP (p \LIMP p) \LIMP p)}$ is 
$\displaystyle {n! \cdot 2 \cdot (\sum_{k=1}^{n+1} k + (n+1) \cdot 2n + \sum_{k=1}^{2n-1} k + 2n)}$, which is equal to $n! \cdot (9n^2 + 9n + 2)$. 
Among them, the number of the non constant functions is $n! \cdot 2 \cdot n$. In Appendix~\ref{classifyApp} the detail is given.}
We can define 
\begin{enumerate}
\item two constant functions that always return 
$\underbar{0}$ or $\underbar{1}$,
\item $n$ projection functions, which
return the value of an argument directly, and 
\item $n$ functions that are the negation of a projection function.
\end{enumerate}
On the other hand, the number of 
all the $n$-argument functions on $\{ \underline{0}, \underline{1} \}$ is $2^{2^n}$.  
Although we only have very limited number of definable functions, 
nevertheless  we can establish a separation result.

\begin{remark}
In the following discussions,
we identify an IIMLL formula with an another IIMLL formula that is different only up to
a permutation: for example,
$p \LIMP (p \LIMP p) \LIMP (p \LIMP p) \LIMP p$ and $(p \LIMP p) \LIMP
p \LIMP (p \LIMP p) \LIMP p$.
If we restrict IIMLL formulas to IIMLL formulas with an order less than 4 and only with
occurrences
of only one atomic formula $p$,
we find that there are only two IIMLL formulas
that have exactly two closed normal IIMLL proof nets, that is,
$p \LIMP (p \LIMP p) \LIMP (p \LIMP p) \LIMP p$ and $p \LIMP p \LIMP (p
\LIMP p \LIMP p) \LIMP p$.
But unlike $p \LIMP (p \LIMP p) \LIMP (p \LIMP p) \LIMP p$,
we can not obtain our separation result by instantiating $p \LIMP p \LIMP (p
\LIMP p \LIMP p) \LIMP p$ for a propositional variable: 
Only two functions are definable by closed IIMLL proof nets of 
${\scriptstyle (p \LIMP p \LIMP (p
\LIMP p \LIMP p) \LIMP p)
\LIMP (p \LIMP p \LIMP (p \LIMP p \LIMP p) \LIMP p)}$, 
that is, $e_3$ and $e_4$ of Table~\ref{oneargumentfuntable}.\footnote{The closed normal proof nets of ${\scriptstyle \overbrace{\scriptstyle (p \LIMP p \LIMP (p
\LIMP p \LIMP p) \LIMP p) \LIMP \ldots \LIMP (p \LIMP p \LIMP (p
\LIMP p \LIMP p) \LIMP p)}^n \LIMP (p \LIMP p \LIMP (p
\LIMP p \LIMP p) \LIMP p)}$ are interesting. We can only define 
parity check functions like 'exclusive or'. 
We can judge whether the number of the occurrences of $1$ (or $0$) of a given sequence with $n$ bits is odd or even by any such a definable function.}
We can not define the two constant functions $e_1$ and $e_2$. 
Without these constant functions, we can not separate 
two closed proof nets of Figure~\ref{basemultiinstant}  
by instantiating $p \LIMP p \LIMP (p
\LIMP p \LIMP p) \LIMP p$ for $p$. 
That is, for any context $C[]$ with ${p \LIMP p \LIMP (p \LIMP p \LIMP p) \LIMP p}^+$ as the conclusion, 
$C[\underbar{0} [{\scriptstyle p \LIMP p \LIMP (p \LIMP p \LIMP p) \LIMP p}/p]]
= C[\underbar{1} [{\scriptstyle p \LIMP p \LIMP (p \LIMP p \LIMP p) \LIMP p}/p]]$.
This is a justification of our choice of $p \LIMP (p \LIMP p) \LIMP (p \LIMP
p) \LIMP p$.
\end{remark}

\begin{table}[htbp]
\begin{center}
\begin{tabular}{|l|l|l|l|}
\hline
$e_1(\underbar{0})=\underbar{0}$ & $e_2(\underbar{0})=\underbar{1}$ & $e_3(\underbar{0})=\underbar{0}$ & $e_4(\underbar{0})=\underbar{1}$ \\
$e_1(\underbar{1})=\underbar{0}$ & $e_2(\underbar{1})=\underbar{1}$ & $e_3(\underbar{1})=\underbar{1}$ & $e_4(\underbar{1})=\underbar{0}$ \\
\hline
\end{tabular}
\caption[all the definable functions 
on  ${\scriptstyle (p \LIMP (p \LIMP p) \LIMP (p \LIMP p) \LIMP p) \LIMP (p \LIMP (p \LIMP p) \LIMP (p \LIMP p) \LIMP p)}$]
{  all the definable functions 
on  ${\scriptstyle (p \LIMP (p \LIMP p) \LIMP (p \LIMP p) \LIMP p) \LIMP (p \LIMP (p \LIMP p) \LIMP (p \LIMP p) \LIMP p)}$
}
\label{oneargumentfuntable}
\end{center}
\end{table}
\begin{table}[htbp]
\begin{center}
\begin{tabular}{|l|l|l|}
\hline
$f_1(\underbar{0},\underbar{0})=\underbar{0}$ & $f_2(\underbar{0},\underbar{0})=\underbar{1}$ & $f_3(\underbar{0},\underbar{0})=\underbar{0}$ \\
$f_1(\underbar{1},\underbar{0})=\underbar{0}$ & $f_2(\underbar{1},\underbar{0})=\underbar{1}$ & $f_3(\underbar{1},\underbar{0})=\underbar{0}$ \\
$f_1(\underbar{0},\underbar{1})=\underbar{0}$ & $f_2(\underbar{0},\underbar{1})=\underbar{1}$ & $f_3(\underbar{0},\underbar{1})=\underbar{1}$ \\
$f_1(\underbar{1},\underbar{1})=\underbar{0}$ & $f_2(\underbar{1},\underbar{1})=\underbar{1}$ & $f_3(\underbar{1},\underbar{1})=\underbar{1}$ \\
\hline
$f_4(\underbar{0},\underbar{0})=\underbar{1}$ & $f_5(\underbar{0},\underbar{0})=\underbar{0}$ & $f_6(\underbar{0},\underbar{0})=\underbar{1}$ \\
$f_4(\underbar{1},\underbar{0})=\underbar{1}$ & $f_5(\underbar{1},\underbar{0})=\underbar{1}$ & $f_6(\underbar{1},\underbar{0})=\underbar{0}$ \\
$f_4(\underbar{0},\underbar{1})=\underbar{0}$ & $f_5(\underbar{0},\underbar{1})=\underbar{0}$ & $f_6(\underbar{0},\underbar{1})=\underbar{1}$ \\
$f_4(\underbar{1},\underbar{1})=\underbar{0}$ & $f_5(\underbar{1},\underbar{1})=\underbar{1}$ & $f_6(\underbar{1},\underbar{1})=\underbar{0}$ \\
\hline
\end{tabular}
\caption[all the definable functions by 112 closed normal proof nets  
on  ${\scriptstyle (p \LIMP (p \LIMP p) \LIMP (p \LIMP p) \LIMP p) \LIMP (p \LIMP (p \LIMP p) \LIMP (p \LIMP p) \LIMP p) \LIMP (p \LIMP (p \LIMP p) \LIMP (p \LIMP p) \LIMP p)}$]
{all the definable functions by 112 closed normal proof nets  
on  ${\scriptstyle (p \LIMP (p \LIMP p) \LIMP (p \LIMP p) \LIMP p) \LIMP (p \LIMP (p \LIMP p) \LIMP (p \LIMP p) \LIMP p) \LIMP (p \LIMP (p \LIMP p) \LIMP (p \LIMP p) \LIMP p)}$}
\label{twoargumentsfuntable}
\end{center}
\end{table}

\subsection{Separation}
The main purpose of the subsection is to prove the following theorem.
\begin{theorem}
\label{separationtheorem}
Let $\Theta_1$ and $\Theta_2$ be IIMLL proof nets with the same conclusion and with an order less than 4 such that $\Theta_1 \neq \Theta_2$. 
Then there is a context $C[]$ such that \\
$C[\Theta_1[{\scriptstyle p \LIMP (p \LIMP p) \LIMP (p \LIMP p) \LIMP p}/p]] = \underline{0}$ 
and $C[\Theta_2[{\scriptstyle p \LIMP (p \LIMP p) \LIMP (p \LIMP p) \LIMP p}/p]] = \underline{1}$.
\end{theorem}
In order to prove the theorem, 
we need some preparations. \\
At first we remark that 
given a closed normal IIMLL proof net $\Theta$ with an order less than 4, 
we can associate a composition $F$ of second order variables $G_1, \ldots G_m$, 
where each $G_i$ ($1 \le i \le m$) occurs in $F$ linearly and 
corresponds to a second order negative formula occurrence 
in the conclusion of $\Theta$
and, the way that $G_1, \ldots, G_m$ compose is determined by the structure of 
$\Theta$ 
(we can easily define $F$ inductively on the depth of $\Theta$). \\
Let $A^-$ be a second order negative IIMLL formula, that is, 
$A$ has the form $\overbrace{p \LIMP \cdots \LIMP p}^n \LIMP p$. 
Then we define $\arity(A)$ as $n$.

\begin{proposition}
\label{assigncontextprop}
Let $\Theta$ be a normal closed IIMLL proof net with an order less than 4,
$A_1^-, \ldots, A_m^-$ be the second order negative formula occurrences 
in the conclusion of $\Theta$, and
$n$ be the number of all the occurrences of $p^-$ in the conclusion of
$\Theta$. 
Moreover, let 
$g_1, \ldots, g_m$ be functions such that
each $g_i \, (1 \le i \le m)$ is definable by
a closed proof net on 
\[ \scriptstyle{\overbrace{\scriptstyle (p \LIMP (p \LIMP p) \LIMP (p \LIMP p) \LIMP p) \LIMP \ldots \LIMP (p \LIMP (p \LIMP p) \LIMP (p \LIMP p) \LIMP p)}^{\arity(A_i)} \LIMP (p \LIMP (p \LIMP p) \LIMP (p \LIMP p) \LIMP p)},   \]
$c_1, \ldots, c_n$ be a sequence of $\{\underbar{0}, \underbar{1}\}$, 
and $f$ be the linear composition of $g_1, \ldots, g_m$ corresponding to $\Theta$.
Then there is a context $C[]$ such that
$C[\Theta[{\scriptstyle  p \LIMP (p \LIMP p) \LIMP (p \LIMP p) \LIMP p}/p]] \to^\ast c$ iff 
$f(c_1, \ldots, c_n) = c$, where $c$ is an element of $\{\underbar{0}, \underbar{1}\}$.
\end{proposition}

\begin{figure}[htbp]
\begin{center}
\includegraphics[scale=.5]{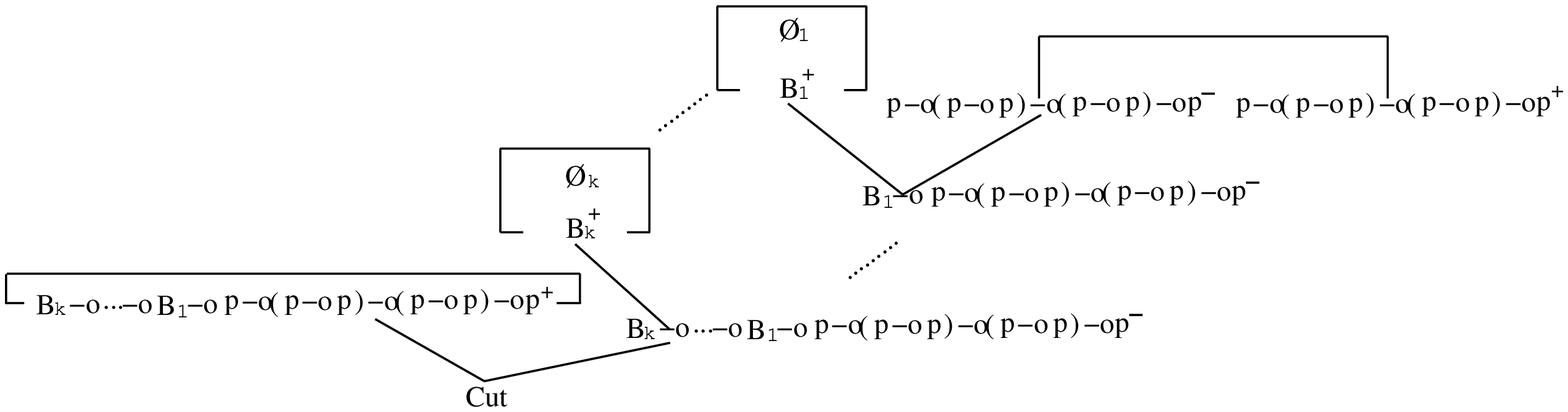}
\caption[a context]
{a context}
\label{assigncontext-iimll}
\end{center}
\end{figure}

\begin{proof}
The conclusion of $\Theta[{\scriptstyle  p \LIMP (p \LIMP p) \LIMP (p \LIMP p) \LIMP p}/p]$ has the form \\
$B_k \LIMP \cdots \LIMP B_1 \LIMP (p \LIMP (p \LIMP p) \LIMP (p \LIMP p) \LIMP p)$.
Moreover, each $B_i \, (1 \le i \le k)$ has 
a closed IIMLL proof net $\Theta_i$ with the conclusion $B_i^+$ 
corresponding to any of $g_1, \ldots, g_m$ or $c_1, \ldots c_n$.
Then we can construct an context $C[]$ shown in Figure~\ref{assigncontext-iimll}. $\Box$
\end{proof}

We note that the construction of $C[]$ only depends on 
the conclusion on $\Theta$, not on $\Theta$ itself.

\begin{flushleft}{\it Proof of Theorem~\ref{separationtheorem}.} \ \ 
We know by Proposition~\ref{assigncontextprop} that 
we can identify a context $C[]$ with an assignment of 
definable functions on $\{ \underbar{0}, \underbar{1} \}$ 
and values in $\{ \underbar{0}, \underbar{1} \}$ 
to the two linear compositions $F_1$ and $F_2$ corresponding to 
$\Theta_1$ and $\Theta_2$.
Since the conclusion of $\Theta_1$ is the same as that of $\Theta_2$, 
$F_1$ and $F_2$ are different expressions such that
\begin{enumerate}
\item [(a)] each variable $x_j \, (1 \le j \le n)$ occurs linearly in both $F_1$ and $F_2$ and 
\item [(b)] each second order variable $G_i \, (1 \le i \le m)$ also occurs linearly in 
both $F_1$ and $F_2$.
\end{enumerate}
We consider the two cases depending on the way 
$\Theta_1$ and $\Theta_2$ differ:
\begin{enumerate}
\item the case where there are $i \, (1 \le i \le m)$ and $j_1$ and  $j_2 \, (1 \le j_1, j_2 \le n)$ 
such that $G_i(\ldots, x_{j_1}, \ldots)$ occurs in $F_1$ and 
$G_i(\ldots, x_{j_2}, \ldots)$ occurs in $F_2$ and $j_1 \neq j_2$, 
where $x_{j_1}$ and $x_{j_2}$ have the same position in $G_i$:\\
Then, there is $G_{i'}$ with the least depth among such $G_i$'s.
Note that the expression
$F_1$ (resp. $F_2$) can be regarded as a tree and
the path from $G_{i'}$ to the root of $F_1$ 
is the same as that of $F_2$.
To each $G_k$ occurrence in the path 
we assign the projection function w.r.t the argument selected by the path. 
To other $G_{k'}$  we assign the constant function that always returns 
$\underbar{0}$.
In addition, we assign $\underbar{0}$ (resp. $\underbar{1}$) 
to $x_{j_1}$  (resp. $x_{j_2}$). 
To other $x_k$ we assign $\underbar{0}$.
Then it is obvious that by the assignment $F_1$ (resp. $F_2$) returns 
$\underbar{0}$ (resp. $\underbar{1}$).

\item otherwise:\\
There is  $i \, (1 \le i \le m)$ such that 
the position of $G_i$ in $F_1$ differs from that of $F_2$.
Then, there is $G_{i'}$ with the least depth among such $G_i$'s in 
$F_1$ or $F_2$.
Without loss of generality, we can assume that $G_{i'}$ in $F_1$ has 
the least depth.
Then to $G_{i'}$ we assign the constant function that always returns 
$\underbar{1}$. 
Again note that the expression
$F_1$ (resp. $F_2$) can be regarded as a tree and
the path from immediately outer $G_{\ell}$ of $G_{i'}$ to the root of $F_1$ 
is the same as that of $F_2$.
To each $G_k$ occurrence in the path 
we assign the projection function w.r.t the argument selected by the path. 
To other $G_{k'}$  we assign the constant function that always returns 
$\underbar{0}$.
To any $x_k$ we assign $\underbar{0}$.
Then it is obvious that by the assignment $F_1$ (resp. $F_2$) returns 
$\underbar{1}$ (resp. $\underbar{0}$). $\Box$
\end{enumerate}
\end{flushleft}

\begin{corollary}[Weak Typed B\"{o}hm Theorem on IIMLL]
\label{weaktypedBoehmTheorem}
Let $\Theta_1$ and $\Theta_2$ be IIMLL proof nets with the same conclusion such that $\Theta_1 \neq \Theta_2$. 
Then there is a context $C[]$ such that 
$C[\Theta_1[{\scriptstyle p \LIMP (p \LIMP p) \LIMP (p \LIMP p) \LIMP p}/p]] = \underline{0}$ 
and $C[\Theta_2[{\scriptstyle p \LIMP (p \LIMP p) \LIMP (p \LIMP p) \LIMP p}/p]] = \underline{1}$.
\end{corollary}

\begin{proof}
By Corollary~\ref{thirdredcontextcor} and Theorem~\ref{separationtheorem}. $\Box$
\end{proof}

\begin{figure}[htbp]
\begin{center}
\includegraphics[scale=.5]{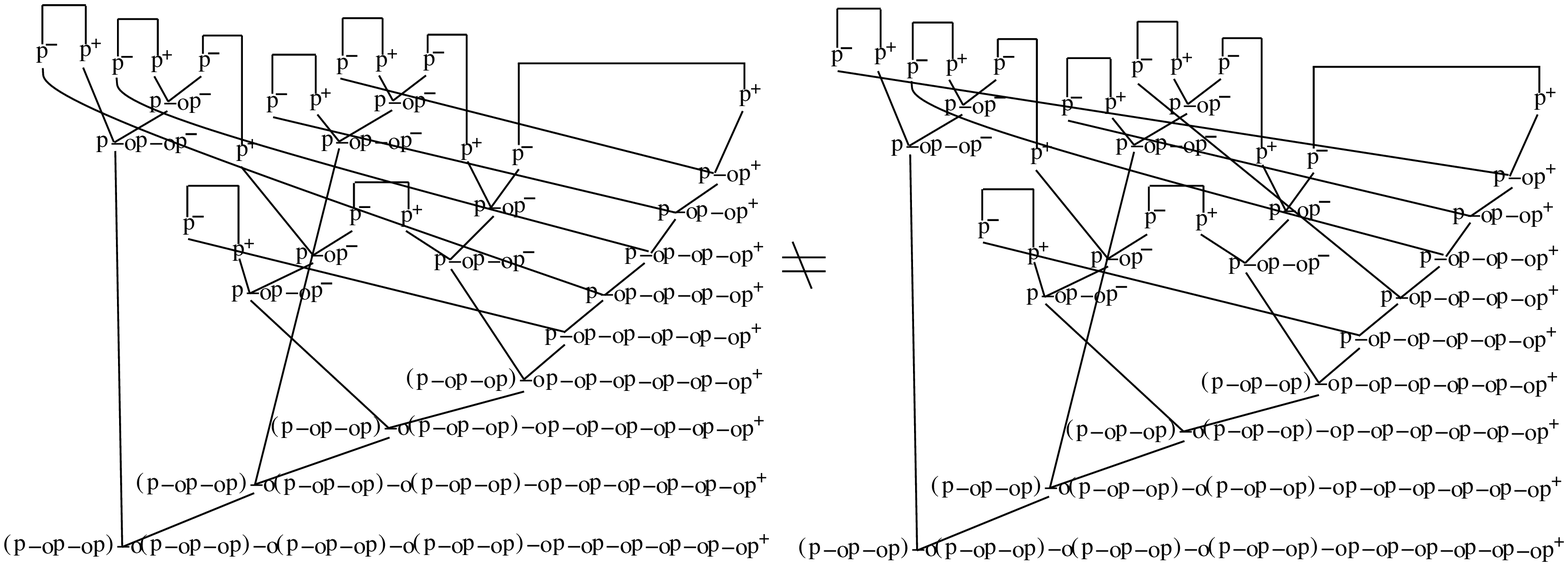}
\caption[an example of the case (1) of the proof of Theorem~\ref{separationtheorem}]
{an example of the case (1) of the proof of Theorem~\ref{separationtheorem}}
\label{case1-ex}
\end{center}
\end{figure}

In the following we explain the proof of Theorem~\ref{separationtheorem} by two examples.
\begin{example}
We explain the case (1) of the proof of Theorem~\ref{separationtheorem}, 
using Figure~\ref{case1-ex}. 
Let $\Theta_1$ (resp. $\Theta_2$) be the left (resp. right) IIMLL proof net of Figure~\ref{case1-ex}.
The expression $F_1$ (resp. $F_2$) corresponding to $\Theta_1$ (resp. $\Theta_2$) is 
$G_1(G_2(x_5, G_4(x_4, x_3)), G_3(x_2,x_1))$ 
(resp. $G_1(G_2(x_5, G_4(x_1, x_3)), G_3(x_2,x_4))$). 
Then we pay attention to the second argument of $G_3$, 
that is, $x_1$ of $F_1$ and $x_4$ of $F_2$, because
the argument in $F_1$ is not the same as that of $F_2$ 
and $G_3$ has the least depth among such second order variables.
Following the proof, we let the context $C[]$ be 
the corresponding to the assignment  
[$x_1 = \underbar{0}$, $x_2=\underbar{0}$, $x_3=\underbar{0}$,
$x_3=\underbar{0}$, $x_4=\underbar{1}$, $x_5=\underbar{0}$,
$G_1=f_3$, $G_2=f_1$, $G_3=f_3$, $G_4=f_1$] (see Table~\ref{twoargumentsfuntable}).
Then $C[\Theta_1] \to^\ast \underbar{0}$ and $C[\Theta_2] \to^\ast \underbar{1}$.
\end{example}

\begin{figure}[htbp]
\begin{center}
\includegraphics[scale=.5]{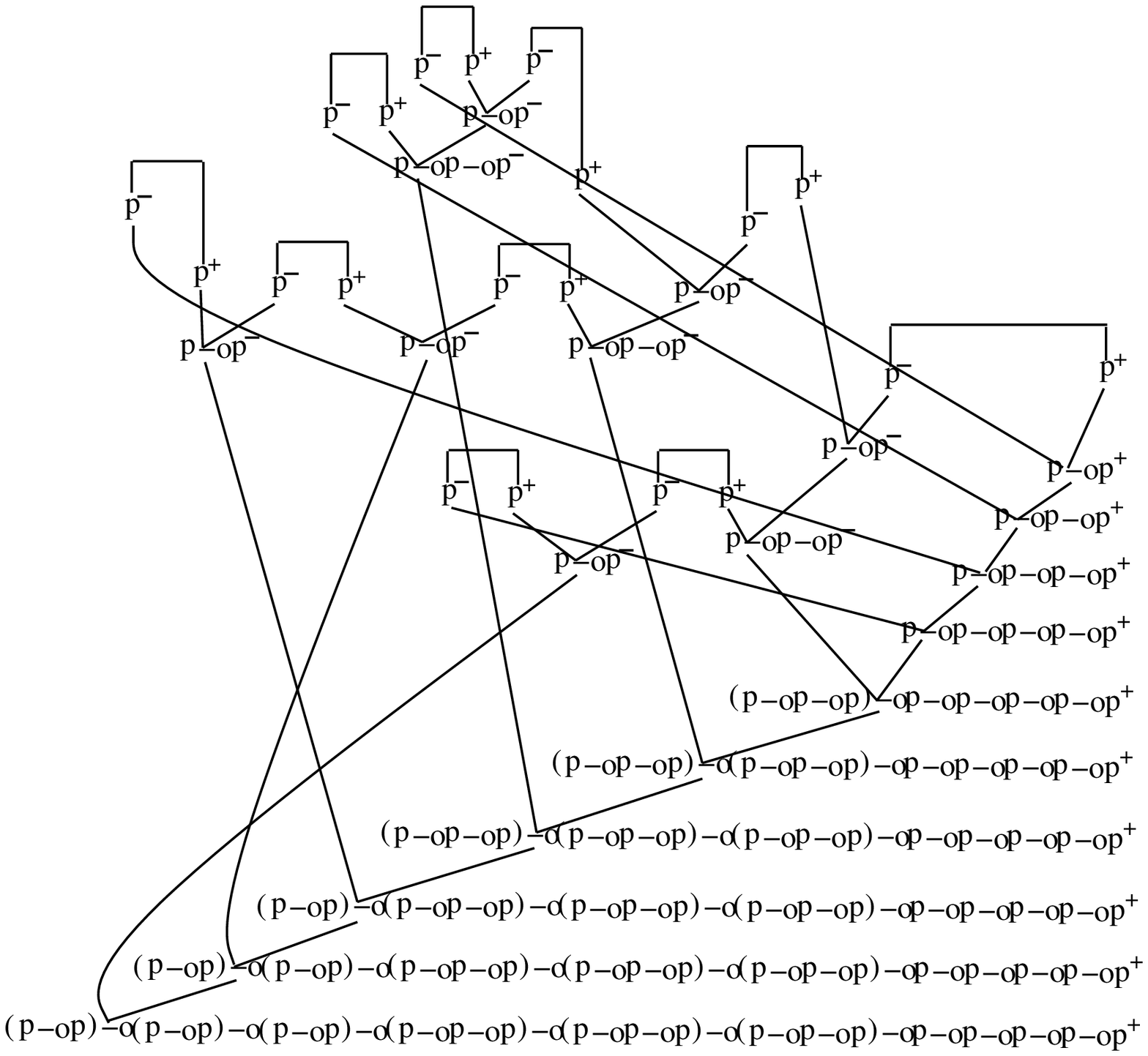}
\caption[an example of the case (2) of the proof of Theorem~\ref{separationtheorem}]{an example of the case (2) of the proof of Theorem~\ref{separationtheorem}}
\label{case2-ex1}
\end{center}
\end{figure}

\begin{figure}[htbp]
\begin{center}
\includegraphics[scale=.5]{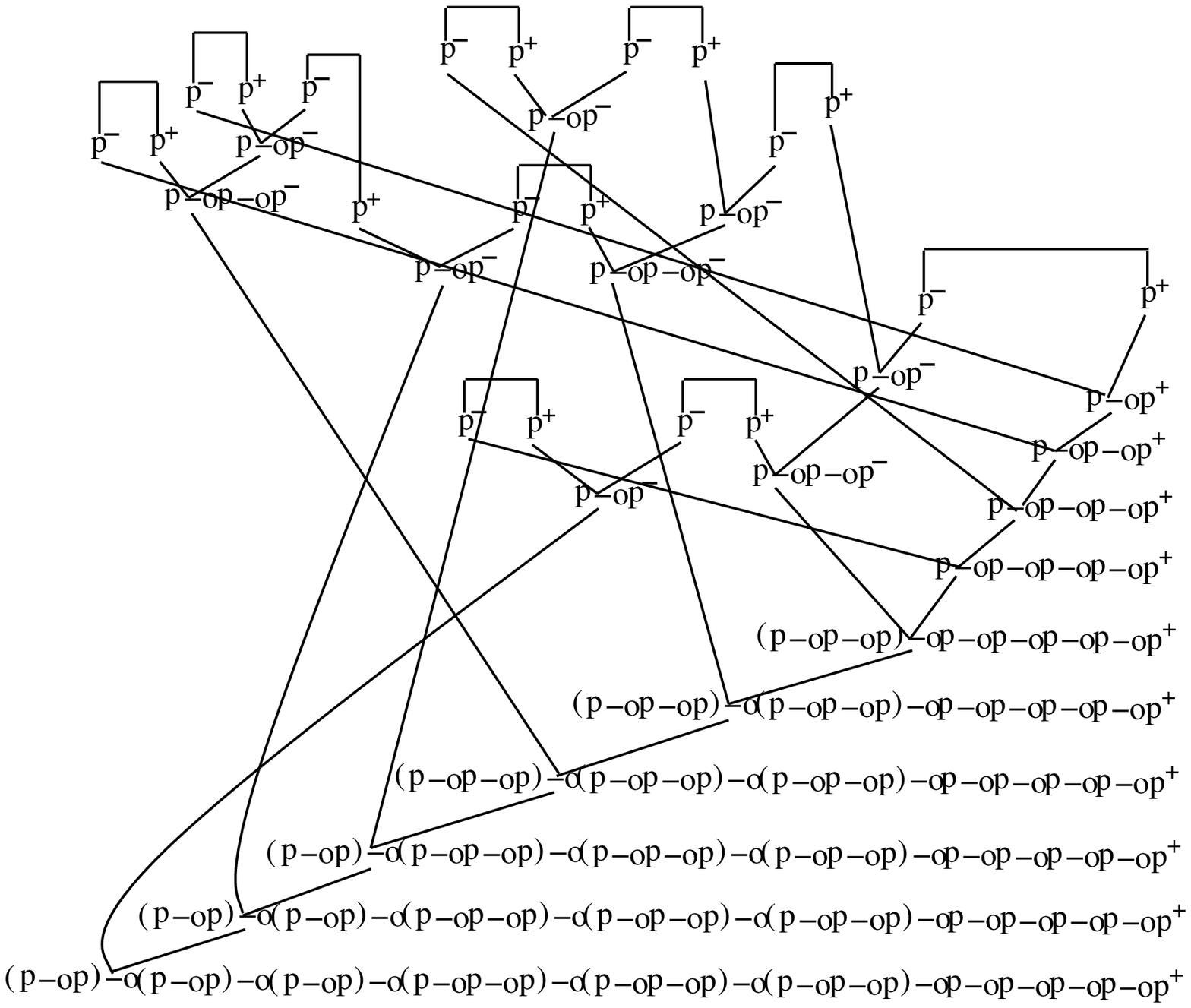}
\caption[an example of the case (2) of the proof of Theorem~\ref{separationtheorem}]{an example of the case (2) of the proof of Theorem~\ref{separationtheorem}}
\label{case2-ex2}
\end{center}
\end{figure}

\begin{example}
We explain the case (2) of the proof of Theorem~\ref{separationtheorem}, 
using Figure~\ref{case2-ex1} and Figure~\ref{case2-ex2}. 
Let $\Theta_1$ (resp. $\Theta_2$) be the IIMLL proof net of Figure~\ref{case2-ex1} (resp. Figure~\ref{case2-ex2}). 
The expression $F_1$ (resp. $F_2$) corresponding to $\Theta_1$ 
(resp. $\Theta_2$) is \\
$G_1(H_3(x_4), G_2(H_2(H_1(x_3)), G_3(x_2,x_1)))$ 
(resp. $G_1(H_3(x_4), G_2(H_2(G_3(x_2,x_1)), H_1(x_3)))$). 
Then we pay attention to the $G_3$ and $H_1$ that have
the position in the second argument of $G_2$ in $F_1$ and $F_2$ respectively,
because the second argument of $G_2$ in $F_1$ are not the same as that 
of $F_2$ and
$G_2$ has the least depth among such second order variables. 
Following the proof, we let the context $C[]$ be 
the corresponding to the assignment  
[$x_1=\underbar{0}$, $x_2=\underbar{0}$, $x_3=\underbar{0}$, $x_4=\underbar{0}$, 
$G_1=f_3$, $G_2=f_3$, $G_3=f_1$, $H_1=e_2$, $H_2=e_1$, $H_3=e_1$ ] 
(see Table~\ref{oneargumentfuntable} and Table~\ref{twoargumentsfuntable}).
Then $C[\Theta_1] \to^\ast \underbar{0}$ and $C[\Theta_2] \to^\ast \underbar{1}$.
\end{example}

\section{An extension to the IMLL case} \label{EXIMLL}
At first we define a special form of third-order IMLL formulas.
\begin{definition}[simple third-order IMLL formulas]
\label{simple}
An IMLL formula $A$ is simple if 
$A$ has the form $\displaystyle B_1 \LIMP \cdots \LIMP B_e \LIMP \overbrace{p \TENS \cdots \TENS p}^{d} \, \, (c \ge 0, d \ge 1, e \ge 0)$, where\\
$\displaystyle B_i = \overbrace{p \TENS \cdots \TENS p}^{k_1} \LIMP \cdots \LIMP \overbrace{p \TENS \cdots \TENS p}^{k_{\ell_i}} \LIMP \overbrace{p \TENS \cdots \TENS p}^{m_i}
\, \, 
(k_j \ge 0, 1 \le j \le \ell_i, \ell_i \ge 1, m_i \ge 1, 1 \le i \le e)$.

\end{definition}

\begin{proposition}
\label{thirdredcontextpropimll}
Let $\Theta_1$ and $\Theta_2$ be closed IMLL proof nets with the same positive conclusion  
such that $\Theta_1 \neq \Theta_2$.
Then there is a context $C[]$ such that $C[\Theta_1] \neq C[\Theta_2]$ and 
the positive conclusion of closed IMLL proof nets $C[\Theta_1]$ and $C[\Theta_2]$ 
is simple.
\end{proposition}

\begin{proof}
Basically the same method as that of Corollary~\ref{thirdredcontextcor}.
$\Box$
\end{proof}

\begin{figure}[htbp]
\begin{center}
\includegraphics[scale=.5]{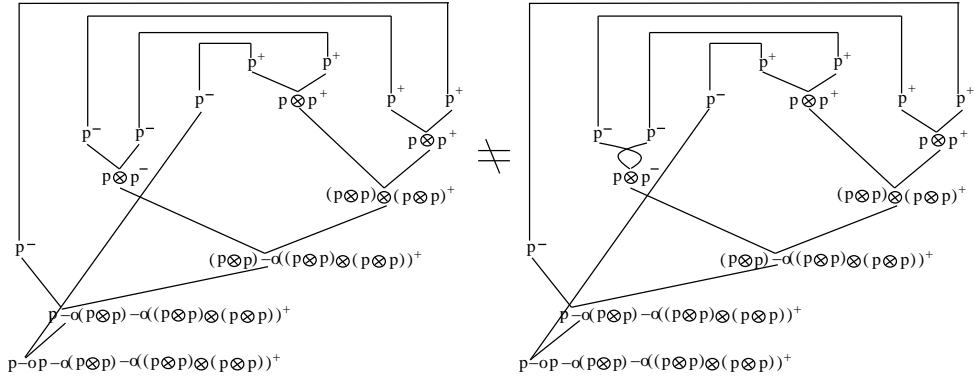}
\caption[two different IMLL proof nets]{two different IMLL proof nets}
\label{reduced-imll-pns}
\end{center}
\end{figure}

For example, the same conclusion of two IMLL proof nets 
of Figure~\ref{imll-pn-ex1} and Figure~\ref{imll-pn-ex2} is not
simple. 
By giving an appropriate context, we can transform these IMLL proof nets 
to two IMLL proof nets with a simple formula as the conclusion in 
Figure~\ref{reduced-imll-pns}.

\begin{proposition}
\label{IIMLLToIMLL}
Let $\Theta_1$ and $\Theta_2$ be closed IMLL proof nets with the same positive simple conclusion 
such that $\Theta_1 \neq \Theta_2$.
Then there is a context $C[]$ such that $C[\Theta_1] \neq C[\Theta_2]$ and 
the positive conclusion of closed IIMLL proof nets $C[\Theta_1]$ and $C[\Theta_2]$ 
has an order less than 4.

\end{proposition}
\begin{proof}
Let the positive simple conclusion of $\Theta_1$ and $\Theta_2$ be $A^+$ in Definition~\ref{simple}.
Then it is obvious to be able to construct 
an IMLL proof net which has conclusions 
$A^-$ and 
$\displaystyle \overbrace{p \LIMP \cdots \LIMP p}^{\sum^e_{i=1} (m_i - 1)} \LIMP 
C_1 \LIMP \cdots \LIMP C_e \LIMP (\overbrace{p \LIMP \cdots \LIMP p}^{d} \LIMP p) \LIMP {p}^{+}$, 
where $\displaystyle C_i=\overbrace{p \LIMP \cdots \LIMP p}^{\sum_{j=1}^{\ell_i} k_j} \LIMP p
\, \, (1 \le i \le e)$. 
It is also obvious to be able to construct a context $C[]$ such that $C[\Theta_1] \neq C[\Theta_2]$ 
and the positive conclusion of $C[\Theta_1]$ and $C[\Theta_2]$ is an intended IIMLL formula. $\Box$
\end{proof}

\begin{figure}[htbp]
\begin{center}
\includegraphics[scale=.5]{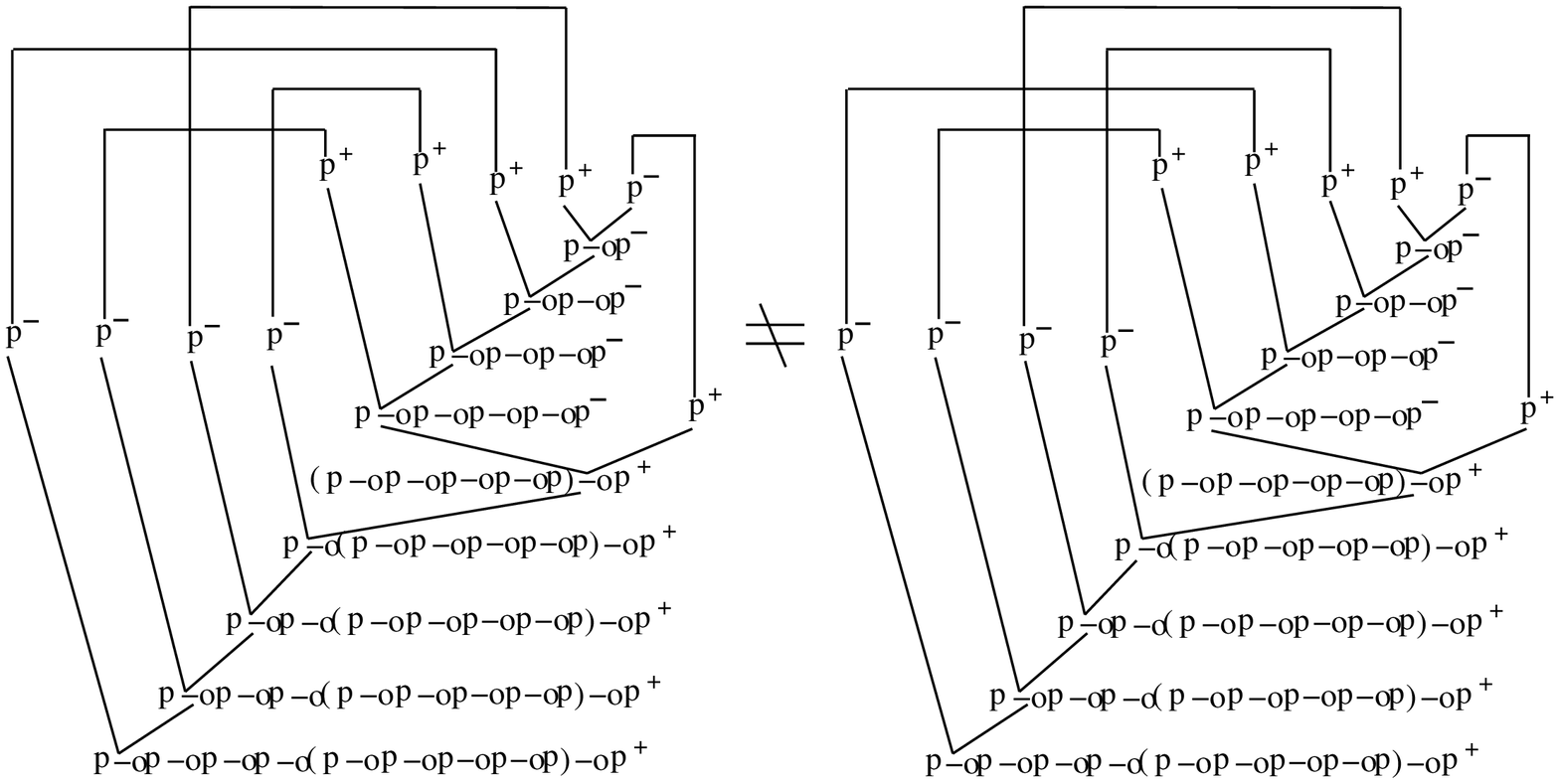}
\caption[two different IIMLL proof nets]{two different IIMLL proof nets}
\label{reduced-iimll-pns}
\end{center}
\end{figure}

For example, there is an IMLL proof net exactly with 
${p_4 \LIMP p_1 \LIMP (p_2 \TENS p_3) \LIMP ((p_5 \TENS p_6) \TENS (p_7 \TENS p_8))}^-$ and 
${p_2 \LIMP p_4 \LIMP p_1 \LIMP p_3 \LIMP (p_5 \LIMP p_6 \LIMP p_7 \LIMP p_8 \LIMP p_0) \LIMP p_0}^+$ as the conclusions, where
the indices of the atomic formula $p$ represent the pairings of ID-links.
From the IMLL proof net, we can construct a context that transforms 
two IMLL proof nets of Figure~\ref{reduced-imll-pns} to
two IIMLL proof nets of Figure~\ref{reduced-iimll-pns}.

From Proposition~\ref{thirdredcontextpropimll}, 
Proposition~\ref{IIMLLToIMLL}, and Theorem~\ref{separationtheorem}, we obtain the following corollary.
\begin{corollary}[Weak Typed B\"{o}hm Theorem on IMLL]
\label{weaktypedBoehmTheoremonIMLL}
Let $\Theta_1$ and $\Theta_2$ be IMLL proof nets with the same conclusion such that $\Theta_1 \neq \Theta_2$. 
Then there is a context $C[]$ such that 
$C[\Theta_1[{\scriptstyle p \LIMP (p \LIMP p) \LIMP (p \LIMP p) \LIMP p}/p]] = \underline{0}$ 
and $C[\Theta_2[{\scriptstyle p \LIMP (p \LIMP p) \LIMP (p \LIMP p) \LIMP p}/p]] = \underline{1}$.
\end{corollary}

\section{Concluding remarks} \label{CONC}
\begin{figure}[htbp]
\begin{center}
\includegraphics[scale=.5]{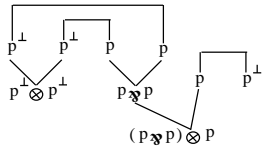}
\caption[A counterexample]{A counterexample}
\label{mll-counter-ex}
\end{center}
\end{figure}

Our result is easily extendable to IMLL {\it with the multiplicative unit} {\bf 1} under a reasonable equality on the extended system,
because the multiplicative unit can be considered as a degenerated IMLL formula.
For example ${\bf 1}^+$ has just one closed proof net and
the closed proof nets on ${{\bf 1} \LIMP p \LIMP (p \LIMP p) \LIMP (p \LIMP p) \LIMP p}^+$ 
have almost the same behaviour as that of 
${p \LIMP (p \LIMP p) \LIMP (p \LIMP p) \LIMP p}^+$. 
However, our separation result w.r.t IMLL with {\bf 1} is stated as follows:
\begin{quote}
Let $\Theta_1$ and $\Theta_2$ be closed IMLL with {\bf 1} proof nets with 
the same positive conclusion such that $\Theta_1 \neq \Theta_2$.
Then there is a context $C[]$ such that $C[\Theta_1]$ and $C[\Theta_2]$
are closed proof nets of ${{\bf 1} \LIMP {\bf 1}}^+$ 
and $C[\Theta_1] \neq C[\Theta_2]$.
\end{quote}
There are two closed normal proof nets of ${{\bf 1} \LIMP {\bf 1}}^+$:
one consists of exactly three links
(an axiom link for ${\bf 1}^+$, a weakening link for ${\bf 1}^-$, 
and a $\PAR$-link). 
Let the proof net be $\underbar{{\tt ff}}_{{{\bf 1} \LIMP {\bf 1}}^+}$.
The other consists of exactly two links
(an ID-link with ${\bf 1}^-$ and ${\bf 1}^+$ and a $\PAR$-link).
Let the proof net be $\underbar{{\tt tt}}_{{{\bf 1} \LIMP {\bf 1}}^+}$.
The proof is similar to that of IMLL without {\bf 1}. \\
However in a symmetric monoidal closed category (SMCC, for example, see \cite{MO03}),
$\underbar{{\tt ff}}_{{{\bf 1} \LIMP {\bf 1}}^+}$ and $\underbar{{\tt tt}}_{{{\bf 1} \LIMP {\bf 1}}^+}$ are interpreted into the same arrow $id_I$, 
where $I$ is the multiplicative unit of a SMCC.
To avoid such an identification, it is possible to relax conditions 
of SMCC: one is to remove the axiom $l_I = r_I$. The other is that
we do not assume $I$ is isomorphic to $I \TENS I$; 
just we assume $I$ is a retract of $I \TENS I$, that is, 
we remove two axioms $l_A ; {l_A}^{-1} = id_{I \TENS A}$
and $r_A ; {r_A}^{-1} = id_{A \TENS I}$. 
The relaxation is quite natural: for example, without these axioms
we can derive important equations like $\alpha_{I,A,B} ; l_{A \TENS B} = 
l_A \TENS id_B$. In the relaxed SMCC, proof nets of IMLL with {\bf 1} can be 
an internal language.
\\
On the other hand, our result cannot be extended to classical multiplicative Linear Logic (for short MLL) directly, 
because all MLL proof nets cannot be polarized by IMLL polarity.
For example, the MLL proof net of Figure~\ref{mll-counter-ex} cannot be transformed 
to an IMLL proof net by type instantiation. \\
As an another direction, fragments including additive connectives may be studied. 
Currently it is proved that our method can be applied to a restricted fragment of intuitionistic 
multiplicative additive linear logic. The restriction is as follows:
\begin{enumerate}
\item With-formulas must positively occur only as $A \WITH A$;
\item Plus-formulas must negatively occur only as $A \PLUS A$.
\end{enumerate}
Moreover we can also prove the strong statement of typed B\"{o}hm theorem 
w.r.t the fragment. 
Our ongoing work is to eliminate the restriction.

\begin{ack}
The author thanks Jean-Jacques Levy, the organizer of the B\"{o}hm theorem workshop at Crete island.
If he had not attended the workshop, he would have not obtain the result. 
He also thanks Martin Hyland, Masahito Hasegawa, Luca Roversi, Alex Simpson, and Izumi Takeuchi 
for helpful comments on the topic. 
\end{ack}

\appendix

\section{A classification}
\label{classifyApp}
\begin{figure}[htbp]
\begin{center}
\includegraphics[scale=.5]{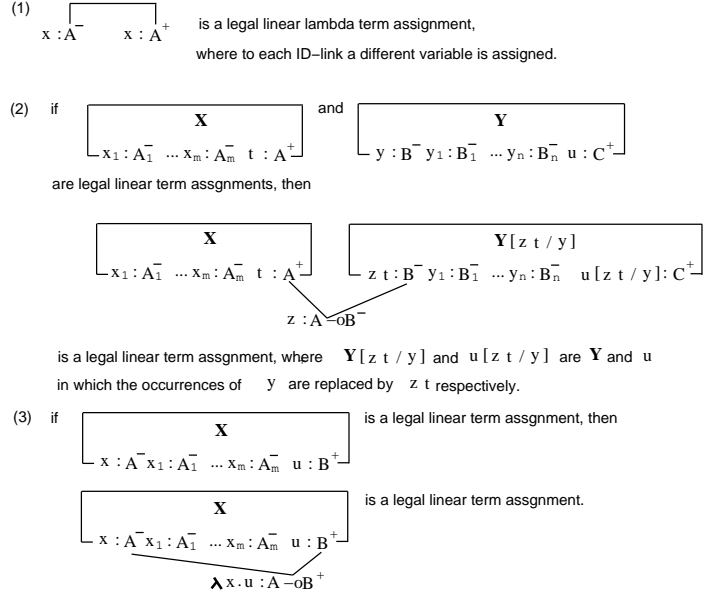}
\caption[A linear $\lambda$-term-assignment system]{A linear $\lambda$-term-assignment system}  
\label{termAssign}
\end{center}
\end{figure}

In this appendix we classify the closed normal IIMLL proof nets of\\ 
${\scriptstyle \overbrace{\scriptstyle (p \LIMP (p \LIMP p) \LIMP (p \LIMP p) \LIMP p) \LIMP \ldots \LIMP (p \LIMP (p \LIMP p) \LIMP (p \LIMP p) \LIMP p)}^{n} \LIMP (p \LIMP (p \LIMP p) \LIMP (p \LIMP p) \LIMP p)}$.
First we introduce a linear $\lambda$-term assignment system to normal IIMLL proof nets, 
since it is easier to discuss the classification in terms of 
$\beta\eta$-long normal linear $\lambda$-terms than 
in terms of normal IIMLL proof nets. 
Figure~\ref{termAssign} shows the term assignment system. 
It is easy to see that all the assigned terms are linear and $\beta\eta$-long normal, 
because to each ID-link with atomic conclusions a different variable is 
assigned and  the first argument in an application term introduced in rule (2) 
is always a variable. 

Second we consider the closed normal linear $\lambda$-terms assigned to ${p \LIMP (p \LIMP p) \LIMP (p \LIMP p) \LIMP p}$. 
While the linear $\lambda$-term $\lambda x.\lambda f.\lambda g. g(f x)$ 
corresponds to the IIMLL proof net $\underbar{0}$, 
$\lambda x.\lambda f.\lambda g. f(g x)$ corresponds to $\underbar{1}$.

\subsection{The closed normal terms on  ${\scriptstyle (p \LIMP (p \LIMP p) \LIMP (p \LIMP p) \LIMP p) \LIMP (p \LIMP (p \LIMP p) \LIMP (p \LIMP p) \LIMP p)}$}
Next we classify the closed $\beta\eta$-long normal terms of \\
${\scriptstyle (p \LIMP (p \LIMP p) \LIMP (p \LIMP p) \LIMP p) \LIMP (p \LIMP (p \LIMP p) \LIMP (p \LIMP p) \LIMP p)}$ as a preliminary step. 
Since the closed $\beta\eta$-long normal terms on the formula have always the form
$\lambda F. \lambda x. \lambda f. \lambda g. t$, 
we only write down the body $t$ instead of writing down the whole term in the following. \\
We classify them according to the surrounding contexts of $f$ and $g$.
\begin{enumerate}
\item [(a)] The case where $\lambda y. f (g y)$ or $\lambda y. g (f y)$ occurs as a subterm:\\
	(1) $F x (\lambda y_1.y_1)(\lambda y_2. f (g y_2))$ and
	(2) $F x (\lambda y_1.y_1)(\lambda y_2. g (f y_2))$ and\\
	(3) $F x (\lambda y_1. f (g y_1))(\lambda y_2.y_2)$ and
	(4) $F x (\lambda y_1. g (f y_1))(\lambda y_2.y_2)$
\item [(b)] The case where both $\lambda y. f y$ and $\lambda y. g y$ occur 
as a subterm:\\
	(5) $F x (\lambda y_1. f y_1) (\lambda y_2. g y_2)$ and
	(6) $F x (\lambda y_1. g y_1) (\lambda y_2. f y_2)$\\
	While the first term denotes the identity function on 
	$\{\underbar{0}, \underbar{1} \}$,
	the second term the negation. 
	The terms of the other cases are a constant function on 
	$\{\underbar{0}, \underbar{1} \}$.
	Note that in order for a term to denote a non-constant function, 
	in the term, $f$ and $g$ must occur 
	in the second argument and the third argument of $F$ separately, 
	because for $F$,  $\lambda x.\lambda f.\lambda g. g(f x)$ or
	$\lambda x.\lambda f.\lambda g. f(g x)$ is substituted.
\item [(c)] The case where $\lambda y. f y$ (respectively $\lambda y. g y$) occurs 
	as a subterm, but $\lambda y. g y$ (respectively $\lambda y. f y$) 
	does not:\\
	(7) $f (F x (\lambda y_1. y_1) (\lambda y_2. g y_2))$ and
	(8) $g (F x (\lambda y_1. y_1) (\lambda y_2. f y_2))$ and\\
	(9) $f (F x (\lambda y_1. g y_1) (\lambda y_2.y_2))$ and
	(10) $g (F x (\lambda y_1. f y_1) (\lambda y_2.y_2))$ and\\
	(11) $F (f x) (\lambda y_1. y_1) (\lambda y_2. g y_2)$ and
	(12) $F (g x) (\lambda y_1. y_1) (\lambda y_2. f y_2)$ and\\
	(13) $F (f x) (\lambda y_1. g y_1) (\lambda y_2. y_2)$ and
	(14) $F (g x) (\lambda y_1. f y_1) (\lambda y_2. y_2)$.
\item [(d)] The case where neither $\lambda y. f (g y)$, $\lambda y. g (f y)$, 
	$\lambda y. f y$, nor $\lambda y. g y$ occurs as a subterm:\\
	(15) $f(g(F x (\lambda y_1. y_1) (\lambda y_2. y_2)))$ and 
	(16) $g(f(F x (\lambda y_1. y_1) (\lambda y_2. y_2)))$ and\\
	(17) $f(F (g x) (\lambda y_1. y_1) (\lambda y_2. y_2))$ and
	(18) $g(F (f x) (\lambda y_1. y_1) (\lambda y_2. y_2))$ and\\
	(19) $F (f (g x)) (\lambda y_1. y_1) (\lambda y_2. y_2)$ and
	(20) $F (g (f x)) (\lambda y_1. y_1) (\lambda y_2. y_2)$
\end{enumerate}
\subsection{The general case}
Finally, we classify the closed normal terms of \\
${\scriptstyle \overbrace{\scriptstyle (p \LIMP (p \LIMP p) \LIMP (p \LIMP p) \LIMP p) \LIMP \ldots \LIMP (p \LIMP (p \LIMP p) \LIMP (p \LIMP p) \LIMP p)}^{n} \LIMP (p \LIMP (p \LIMP p) \LIMP (p \LIMP p) \LIMP p)}$. Since the closed $\beta\eta$-long normal terms on the formula has always the form
$\lambda F_1.\cdots \lambda F_n. \lambda x. \lambda f. \lambda g. t$, 
we only write down the body $t$ instead of writing down the whole term in the following. \\
The classification proceeds in the same fashion as that of the previous subsection:
\begin{enumerate}
\item [(a)] The case where $\lambda y. f (g y)$ or $\lambda y. g (f y)$ occurs as a subterm:\\
	In this case, $t$ has the form 
	\[ F_1(\cdots (F_{n-1} (F_n \, \, x \, \, t_{2n-1} t_{2n}) t_{2n-3} t_{2n-2}) \cdots) t_1 t_2\] or a permutation on $\{ F_1,\ldots, F_n \}$ of the form, 
	where $t_i \, (1 \le i \le 2n)$ is $\lambda y. f (g y)$, $\lambda y. g (f y)$ or $\lambda y.y$, but 
any of $\lambda y. f (g y)$ and $\lambda y. g (f y)$ exclusively occurs once.
The total number of such terms is $n! \times 2 \times 2 n$.
\item [(b)] The case where both $\lambda y. f y$ and $\lambda y. g y$ occur 
as a subterm:\\
	In this case, $t$ has the form 
	\[ F_1(\cdots (F_{n-1} (F_n \, \, x \, \, t_{2n-1} t_{2n}) t_{2n-3} t_{2n-2}) \cdots) t_1 t_2\] or a permutation on $\{ F_1,\ldots, F_n \}$ of the form, 
	where $t_i \, (1 \le i \le 2n)$ is $\lambda y. f y$, $\lambda y. g y$ or $\lambda y.y$, and both $\lambda y. f y$ and $\lambda y. g y$ occur exactly once.
The total number of such terms is 
$n! \times 2 \times {\scriptstyle {2n}} C_2 
= n! \times 2 \times \sum^{2n-1}_{k=1} k= n! \times 2 \times (2n^2-n)$. 
Among such terms the total number of the terms in which 
there is an $i \, (1 \le i \le n)$ such that 
both the second argument and the third argument of $F_i$ 
are exactly $\lambda y. f y$ or $\lambda y. g y$ 
is $n! \times 2 \times n$. 
Only such limited terms are 
a non-constant function, i.e., a projection or the negation of such a projection.
Other terms of the case and the terms of the other cases are a constant function.
\item [(c)] The case where $\lambda y. f y$ (respectively $\lambda y. g y$) occurs 
	as a subterm, but $\lambda y. g y$ (respectively $\lambda y. f y$) 
	does not:\\
	In this case, $t$ has the form 
	\[ 
	h_1 (F_1 (h_2 (F_2 ( \cdots ( h_{n-1} (F_{n-1} (h_n 
	(F_n (h_{n+1} x) t_{2n-1} t_{2n})) t_{2n-3} t_{2n-2})) \cdots ) t_3 t_4))
	t_2 t_1)
	\]
	or a permutation on $\{ F_1,\ldots, F_n \}$ of the form, where $h_i \, (1 \le i \le n+1)$ is empty or $g$ (resp. $f$), and
	$g$ (resp. $f$) occurs exactly once. 
	Moreover $t_j \, (1 \le j \le 2n)$ is 
	$\lambda y.y$ or $\lambda y.f y$  (resp. $\lambda y.g y$) and 
	$\lambda y.f y$  (resp. $\lambda y.g y$) occurs exactly once.
	The total number of such terms is $n! \times 2 \times ((n+1) \times 2n)$.
\item [(d)] The case where neither $\lambda y. f (g y)$, $\lambda y. g (f y)$, 
	$\lambda y. f y$, nor $\lambda y. g y$ occurs as a subterm:\\
	In this case, $t$ has the form 
	\[ 
	h_1 (F_1 (h_2 (F_2 ( \cdots ( h_{n-1} (F_{n-1} (h_n 
	(F_n (h_{n+1} x) t_{2n-1} t_{2n})) t_{2n-3} t_{2n-2})) \cdots ) t_3 t_4))
	t_2 t_1)
	\]
	or a permutation on $\{ F_1,\ldots, F_n \}$ of the form, where $h_i \, (1 \le i \le n+1)$ is empty, $f$, $g$, $f (g[])$, or
	$g (f [])$, and both $f$ and $g$ occur exactly once.
	Moreover $t_j \, (1 \le i \le 2n)$ is always $\lambda y.y$.
	The total number of such terms is 
	$n! \times 2 \times \sum^{n+1}_{k=1} k = n! \times (n^2 + 3n +2)$.

\end{enumerate}
\end{document}